\documentclass[aps,twocolumn,pra,superscript,floatfix,superscriptaddress,showpacs,footinbib]{revtex4-1}
\usepackage{amssymb}

\usepackage[pdftex]{graphicx}
\usepackage{dcolumn}
\usepackage{bm}
\usepackage{amsmath}
\usepackage{array}
\usepackage{color}
\usepackage{float}
\usepackage{subfigure}
\usepackage{dsfont}
\usepackage{xcolor}
\usepackage{bbold}

\newcommand{\+}{\dagger}

\newcommand{\e}{\varepsilon}

\newcommand{\s}{\sigma}

\newcommand{\up}{\uparrow}
\newcommand{\down}{\downarrow}

\newcommand{\veck}{\mathbf{k}}

\newcommand\smallO{
  \mathchoice
    {{\scriptstyle\mathcal{O}}}
    {{\scriptstyle\mathcal{O}}}
    {{\scriptscriptstyle\mathcal{O}}}
    {\scalebox{.6}{$\scriptscriptstyle\mathcal{O}$}}
  }

\usepackage{hyperref}
\hypersetup{
    colorlinks,%
    citecolor=blue,%
    linkcolor=blue,%
    urlcolor=blue
}

\begin{document}

\title{Multi-Dirac and Weyl physics in heavy-fermion systems}

\author{Joelson F. Silva}
\affiliation{Gleb Wataghin Institute of Physics, The University of Campinas (Unicamp), 13083-859 Campinas, SP, Brazil}

\author{E. Miranda}
\affiliation{Gleb Wataghin Institute of Physics, The University of Campinas (Unicamp), 13083-859 Campinas, SP, Brazil}


\begin{abstract}
We have studied multi-Dirac/Weyl systems with arbitrary topological charge $n$ in the presence of a lattice of local magnetic moments. To do so, we propose a multi-Dirac/Weyl Kondo lattice model which is analyzed through a mean-field approach appropriate to the paramagnetic phase. We study both the broken time-reversal and the broken inversion-symmetry Weyl cases. The multi-Dirac and broken-time reversal multi-Weyl cases have similar behavior, which is in contrast to the broken-parity case. For the former, low-energy particle-hole symmetry leads to the emergence of a critical coupling constant below which there is no Kondo quenching, reminiscent of the pseudogap Kondo impurity problem. Away from particle-hole symmetry, there is always Kondo quenching. For the broken inversion symmetry, there is no critical coupling. Depending on the conduction electron filling, Kondo insulator, heavy fermion metal or semimetal phases can be realized. In the last two cases, quasiparticle renormalizations can differ widely between opposite chirality sectors, with characteristic dependences on  microscopic parameters that could in principle be detected experimentally. 
\end{abstract}
\maketitle

\section{Introduction}

In the last several years much attention has been devoted to systems that show non-trivial topological states of matter~\cite{Hasan2010,Moore2010}. The main examples of topological systems are topological insulators and topological superconductors~\cite{Bernevig2013,Sato2017}. Besides their fascinating physics, this focus is also due to some of their characteristic features that can find applications in a wide range of areas, from electronic transport to quantum computation~\cite{Wilczek2006,Nayak2008,Tian2017}.

Other important examples of topological systems are topological semimetals~\cite{Burkov2016}. When time-reversal and inversion symmetries are present, topological semimetals are characterized by degenerate 3D Dirac cones. If, on the other hand, time-reversal symmetry~(TRS) or inversion symmetry~(IS) is broken, the Dirac nodes split into Weyl nodes and the 
system becomes a Weyl semimetal~\cite{Xu2015,doi:10.1146/annurev-conmatphys-031016-025458}. Weyl nodes act as monopoles of 3D Bery curvature characterized by the topological charge $n$. 
In 3D translationally invariant systems only topological charges $n=1,2$ and $3$ are allowed~\cite{PhysRevLett.108.266802}.  
Most known Weyl semimetals are single-Weyl semimetals, characterized by unitary topological charge~($n=1$)~\cite{RevModPhys.90.015001}. However, 
double-Weyl nodes~($n=2$) are predicted to occur, for example, in ${\rm HgCr}_{2}{\rm Se}_{4}$ and ${\rm SrSi}_{2}$~\cite{PhysRevLett.108.266802,Huang2016}. Moreover, triple-Weyl nodes~($n=3$) can be found in transition-metal monochalcogenides ${\rm A(MoX)}_{3}$, with ${\rm A}={\rm Rb},{\rm Tl}$; ${\rm X}={\rm Te}$~\cite{PhysRevX.7.021019}. 

The presence of single-Weyl nodes in the heavy fermion 
compound ${\rm Ce}_3{\rm Bi}_4{\rm Pd}_3$~\cite{PhysRevLett.118.246601} has been proposed as an explanation of some its exotic physical properties, such as a giant spontaneous Hall effect in a nonmagnetic regime~\cite{doi:10.1073/pnas.1715851115,PhysRevB.101.075138,Dzsaber2021}. These systems are referred to as \textit{Weyl-Kondo semimetals}~\cite{doi:10.1073/pnas.1715851115}. Furthermore, the presence of double-Weyl nodes has been predicted in Weyl semimetals without inversion symmetry~\cite{PhysRevB.99.035141}. Finally, the interest in Dirac(Weyl)-Kondo semimetals goes beyond the condensed matter community, as they are a platform for the study of relativistic fermions in the presence of quantum impurities, such  as the ``QCD Kondo effect'', which may be realized in quark matter systems~\cite{PhysRevC.88.015201,PhysRevD.92.065003,PhysRevResearch.2.023312,PhysRevD.104.094515}.  

Despite the existence of previous studies of multi-Dirac(Weyl) systems in the presence of single and double magnetic impurities~\cite{PhysRevB.92.041107,PhysRevB.92.121109,Sun2017,PhysRevB.99.115109,PhysRevB.103.045137}, less attention has been devoted to multi-Dirac(Weyl) semimetals in the presence of an ordered lattice of magnetic moments~\cite{PhysRevLett.118.246601,doi:10.1073/pnas.1715851115,PhysRevB.101.075138,Dzsaber2021}. It is the aim of this paper address this question. We propose here the study of a multi-Weyl Kondo lattice model with arbitrary topological charge $n$. As a first step in that direction, we focus on the paramagnetic phase of the model, where a large-$N$ inspired mean-field approach has proved extremely useful for the understanding of experiments in the usual Kondo lattice case. Prominent among the latter are the interaction-induced renormalizations that lead to the large effective masses in heavy fermion metals and the reduced hybridization gaps in Kondo insulators. In the multi-Dirac(Weyl) Kondo lattice, we find three possible phases: the multi-Dirac(Weyl) Kondo insulator, the heavy fermion semimetal and the heavy fermion metal. We distinguish, on the one hand, the multi-Dirac and the broken-TRS multi-Weyl Kondo lattices, both of which have similar properties, from the broken-IS multi-Weyl case, on the other hand. Our main results are a thorough characterization of the different renormalizations of Kondo gaps, quasiparticle masses and velocities. These will depend on whether particle-hole symmetry is present or not. In particular, in the broken-IS multi-Weyl Kondo lattice, different chirality sectors will suffer vastly different renormalizations, which should lead to discernible experimental signatures that we will discuss.

The paper is organized as follows. In Sec.~\ref{sec:model} we present the multi-Dirac(Weyl) Kondo lattice model. In Sec.~\ref{sec:mean-field} we describe the general mean-field approach used. In Sec.~\ref{sec:multi-Dirac Kondo} we present the results of this approach as applied to the multi-Dirac and multi-Weyl Kondo systems. We analyze the mean-field results in Sec.~\ref{sec:discussion} in light of some previous theoretical results and point to the measurable signatures of our findings.  Finally, in Sec.~\ref{sec:conclusions}, we wrap  up  with some concluding remarks. Some mathematical developments are left to the Appendices.

\section{Model}\label{sec:model}
We focus on a model of a multi-Weyl conduction band coupled to a lattice of quantum spins, aptly described as a multi-Weyl-Kondo 
lattice. The Hamiltonian consists then of two terms 
\begin{equation}\label{H_model}
 H=H_{0}+H_{K},
\end{equation}
where $H_{0}$ corresponds to the multi-Weyl semimetal 
described in $\veck$-space by the minimal model~\cite{PhysRevB.95.201102,PhysRevB.97.045150,PhysRevB.99.115109}  
\begin{equation}\label{h_0}
H_{0}=\sum_{\veck}\Psi_{\veck}^{\+}\mathcal{H}_{\veck}\Psi_{\veck}-\mu,
\end{equation}
where $\Psi_{\veck}=(c_{\veck +\up},c_{\veck +\down},c_{\veck -\up},c_{\veck -\down})^{T}$,
$c_{\veck s\alpha}^{\+}$ is 
the creation operator for an electron in state $\veck$ of orbital $s$ and spin $\alpha$,
$\mu$ is chemical potential,
\begin{eqnarray}\label{h_kk}
 \mathcal{H}_{\veck}&=&\tau_{z}\otimes[v_{\perp}k_{0}^{(1-n)}({k}_{-}^{n}\sigma_{+}+{k}_{+}^{n}\sigma_{-})+v_{z}k_{z}\sigma_{z}-Q_{0}\sigma_{0}] \nonumber \\
 &&-v_{z}Q\tau_{0}\otimes \sigma_{z},
\end{eqnarray}
$k_{0}$ is a dimensionful reference wave vector, $k_{\pm}=k_{x}\pm i k_{y}$, $\sigma_{\pm}=(\sigma_{x}\pm i\sigma_{y})/2$, and $n=1,2,3$ is the topological charge that characterizes the multi-Weyl semimetal index. Besides, $\boldsymbol{\sigma}=(\sigma_{x},\sigma_{y},\sigma_{z})$ are the 
Pauli matrices which act on the (pseudo)spin-space, $\boldsymbol{\tau}=(\tau_{x},\tau_{y},\tau_{z})$ act 
on the orbital space, and $\sigma_{0}$ and $\tau_{0}$ denote the identity matrices in spin and orbital spaces, respectively. The term in $Q$ breaks time reversal symmetry~(TRS), since 
for $Q=0$ we have that $\mathcal{T}\mathcal{H}_{\veck}(\veck)\mathcal{T}^{-1}=\mathcal{H}_{\veck}(-\veck)$, where $\mathcal{T}=\tau_{0}\otimes(i\sigma_{y})K$ 
is the time reversal operator (with complex conjugation $K$). On the other hand, 
the term in $Q_{0}$ breaks inversion symmetry~(IS). For $Q_{0}=0$, $\mathcal{P}\mathcal{H}_{\veck}(\veck)\mathcal{P}^{-1}=\mathcal{H}_{\veck}(-\veck)$, where $\mathcal{P}=\tau_{x}\otimes\sigma_{0}$ 
is the inversion operator.
When both time reversal and inversion symmetries are present~($Q=Q_{0}=0$) the model reduces to a 3D multi-Dirac semimetal~\cite{doi:10.1126/science.1245085,Neupane2014}. A generic Kondo Hamiltonian $H_{K}$ has two kinds of contributions, associated with 
intra-orbital processes and inter-orbital processes~\cite{PhysRevB.71.115312,PhysRevLett.95.067204}. Thus, it is given by 
\begin{eqnarray}\label{H_Ks}
H_{K}&=&\frac{1}{2}\sum_{js,\alpha,\beta}(J\mathbf{S}_{js}+W\mathbf{S}_{j\bar{s}})\cdot (c_{js\alpha}^{\+}\boldsymbol{\sigma}_{\alpha\beta}c_{js\beta}) \nonumber \\
 &&+\frac{1}{2}\sum_{js,\alpha,\beta}K(\mathbf{S}_{js}+\mathbf{S}_{j\bar{s}})\cdot\left[(c_{js\alpha}^{\+}\boldsymbol{\sigma}_{\alpha\beta}c_{j\bar{s}\beta}) \right. \nonumber \\
 &&\left.\qquad+(c_{j\bar{s}\alpha}^{\+}\boldsymbol{\sigma}_{\alpha\beta}c_{js\beta})\right] \nonumber \\
\end{eqnarray}
where $\alpha,\beta$ are spin components, $s=\pm$ are the orbital indices, $\bar{s}=-s$ , $c_{js\alpha}^{\+}=\frac{1}{\sqrt{\mathcal{V}}}\sum_{\veck}c_{\veck s\alpha}^{\+}e^{-i\veck\cdot \mathbf{R}_{j}}$, and $\cal{V}$ is the number of unit cells. The first term of $H_{K}$ is associated with intra-orbital processes, while the second refers to inter-orbital ones, both mediated by the interaction with the local magnetic moments. The latter are assumed to have, in general, orbital indices as well.

\section{ Mean-field Approach}\label{sec:mean-field}
\noindent
The local magnetic moment operators that appear in~\eqref{H_Ks} can be written using Abrikosov's pseudo-fermion operators $f_{js\alpha}$ with both spin and orbital indices
as
\begin{equation}
\mathbf{S}_{js}=\frac{1}{2}\sum_{\alpha\beta}
f_{js\alpha}^{\+}\boldsymbol{\sigma}_{\alpha\beta}f_{js\beta}
\end{equation}
A faithful representation of spin-1/2 operators requires the enforcement of the single-occupancy constraint (per site and orbital)
\begin{equation}\label{constraint}
\sum_{\alpha}
f_{js\alpha}^{\+}f_{js\alpha}=n_{fjs}=1.
\end{equation}
Using this we can write $H_{K}$ up to constant factors as 
\begin{eqnarray}\label{H_field}
H_K&=&
 -\frac{J}{2}\sum_{js,\alpha,\beta}(c_{js\beta}^{\+}f_{js\beta})(f_{js\alpha}^{\+}c_{js\alpha}) \nonumber \\
 &&-\frac{W}{2}\sum_{js,\alpha,\beta}(c_{j\bar{s}\beta}^{\+}f_{j{s}\beta})(f_{js\alpha}^{\+}c_{j\bar{s}\alpha}) \nonumber \\
 &&-\frac{K}{2}\sum_{js,\alpha,\beta}\left[(c_{j+\beta}^{\+}f_{js\beta})(f_{js\alpha}^{\+}c_{j-\alpha})+{\rm H.c}\right] \nonumber \\
 &&\qquad+\tilde{\lambda}\sum_{js}(n_{fjs}-1).
\end{eqnarray}
In the last line, we introduced the Lagrange multiplier $\tilde{\lambda}$ anticipating the later enforcement of the constraint $n_{fjs}=1$ at the mean field level. 

\subsection{Mean-field Hamiltonian and Free energy}
\noindent 
We now proceed along the lines of the large-$N$-inspired mean-field treatment of Kondo lattices~\cite{Read1983,Coleman1983,Auerbach1986}.
We do so in Hamiltonian language through the following decouplings of the quartic terms in the usual Kondo channels
\begin{eqnarray*}
(c_{js\beta}^{\+}f_{js\beta})(f_{js\alpha}^{\+}c_{js\alpha})&\rightarrow & 
\langle c_{js\beta}^{\+}f_{js\beta}\rangle f_{js\alpha}^{\+}c_{js\alpha} \\
&&+\langle f_{js\alpha}^{\+}c_{js\alpha}\rangle c_{js\beta}^{\+}f_{js\beta} \\
&&- \langle c_{js\beta}^{\+}f_{js\beta}\rangle \langle f_{js\alpha}^{\+}c_{js\alpha}\rangle , \\
(c_{j\bar{s}\beta}^{\+}f_{j{s}\beta})(f_{js\alpha}^{\+}c_{j\bar{s}\alpha})&\rightarrow &\langle c_{j\bar{s}\beta}^{\+}f_{j{s}\beta} \rangle f_{js\alpha}^{\+}c_{j\bar{s}\alpha} \\
&&+\langle f_{js\alpha}^{\+}c_{j\bar{s}\alpha}\rangle c_{j\bar{s}\beta}^{\+}f_{j{s}\beta} \\
&& -\langle c_{j\bar{s}\beta}^{\+}f_{j{s}\beta} \rangle \langle f_{js\alpha}^{\+}c_{j\bar{s}\alpha}\rangle , \\
(c_{j+\beta}^{\+}f_{js\beta})(f_{js\alpha}^{\+}c_{j-\alpha})&\rightarrow & \langle c_{j+\beta}^{\+}f_{js\beta}\rangle f_{js\alpha}^{\+}c_{j-\alpha} \\
&&+\langle f_{js\alpha}^{\+}c_{j-\alpha} \rangle c_{j+\beta}^{\+}f_{js\beta} \\
&&-\langle c_{j+\beta}^{\+}f_{js\beta}\rangle \langle f_{js\alpha}^{\+}c_{j-\alpha} \rangle .
\end{eqnarray*}
Defining the spinors $\Psi_{\veck s}^{\+}=(c_{\veck s\up}^{\dag},c_{\veck s\down}^{\+})$ and $\Psi_{\veck fs}^{\+}=(f_{\veck s\up}^{\+},f_{\veck s\down}^{\+})$, where 
$f_{\veck s \s}^{\+}=\frac{1}{\sqrt{\mathcal{V}}}\sum_{j}f_{js\s}^{\+}e^{i\veck\cdot \mathbf{R}_{j}}$ we can write the mean-field Hamiltonian in the terms of
$\Phi_{\veck}^{\+}=(\Psi_{\veck +}^{\+},\Psi_{\veck f+}^{\+},\Psi_{\veck-}^{\+},\Psi_{\veck f-}^{\+})$ as 
\begin{widetext}
\begin{equation}
H_{MF}=\sum_{\veck}\Phi_{\veck}^{\+}H_{\veck}^{HS}\Phi_{\veck}+\mathcal{V}\left[\sum_{s}(J+K)|v_{s}|^{2}+\sum_{s}(W+K)|w_{s}|^{2}-\lambda  \right],  
\end{equation}
with $\lambda=2\tilde{\lambda}$, enforcing the constraint of Eq.~\eqref{constraint} on the average, and

 \begin{eqnarray}\label{H_HS}
 H_{\veck}^{HS}=\left(\begin{array}{cccc}
 H_{\veck +}-\mu\mathbb{1} & J\mathbb{v}^{+}_{2\times 2}+K\mathbb{w}^{-}_{2\times2} & \mathbb{0}_{2\times 2} & W\mathbb{w}^{+}_{2\times 2}+K\mathbb{v}^{-}_{2\times 2} \\
J{\mathbb{v}}^{*+}_{2\times 2}+K{\mathbb{w}}^{*-}_{2\times2} & \boldsymbol{\lambda}_{2\times 2} & W{\mathbb{w}}^{*+}_{2\times 2}+K{\mathbb{v}}^{*-}_{2\times 2} & \mathbb{0}_{2\times 2} \\
\mathbb{0}_{2\times 2} & W\mathbb{w}^{-}_{2\times2}+K\mathbb{v}^{+}_{2\times 2} & H_{\veck -}-\mu\mathbb{1} & J\mathbb{v}^{-}_{2\times 2}+K\mathbb{w}^{+}_{2\times 2} \\
W{\mathbb{w}}^{*-}_{2\times2}+K{\mathbb{v}}^{*+}_{2\times 2} & \mathbb{0}_{2\times 2} & J{\mathbb{v}}^{*-}_{2\times 2}+K{\mathbb{w}}^{*+}_{2\times 2} & \boldsymbol{\lambda}_{2\times 2}
\end {array}\right).
\end{eqnarray}
\end{widetext}
Above
\begin{eqnarray}
 H_{\veck s}&=&s[v_{\perp}k_{0}^{(1-n)}({k}_{-}^{n}\sigma_{+}+{k}_{+}^{n}\sigma_{-})+v_{z}k_{z}\sigma_{z}-Q_{0}\sigma_{0}] \nonumber \\
 &&\qquad-v_{z}Q \sigma_{z}, \nonumber \\
 \end{eqnarray}
 and the matrices are given by
 \begin{eqnarray}
 \mathbb{v}^{s}_{2\times 2}&=&\left(\begin{array}{cc}
        v_{s} & 0 \\
        0 & v_{s}
       \end{array}\right),\quad \mathbb{w}^{s}_{2\times 2}=\left(\begin{array}{cc}
       w_{s} & 0 \\
       0 & w_{s}
       \end{array} 
\right), \nonumber
\\
\boldsymbol{\lambda}_{2\times 2}&=&\left(\begin{array}{cc}
                                      \lambda & 0 \\
                                      0 & \lambda
                                     \end{array}
\right), \quad
\mathbb{0}_{2\times 2} = \left(\begin{array}{cc}
        0 & 0 \\
        0 & 0
       \end{array}\right),
\end{eqnarray}
where we defined the Kondo order parameters
 \begin{eqnarray}
 v_{s}=\langle c_{js\s}^{\+}f_{js\s}\rangle,\quad w_{s}=\langle c_{js\s}^{\+}f_{j\bar{s}\s}\rangle.
\end{eqnarray}
The equilibrium values of the order parameters are obtained by minimization of the mean-field free-energy density with respect to their values
\begin{eqnarray}\label{F_MF}
 \mathcal{F}_{MF}&=&-\frac{1}{\beta}\sum_{i\omega_{n}}\int\frac{d^{3}\veck}{(2\pi)^{3}}{\rm ln} {\rm Det}[-i\omega_{n}\mathbb{1}+H_{\veck}^{HS}] \nonumber \\
 &&+\left[\sum_{s}(J+K)|v_{s}|^{2} \right. \nonumber \\
 &&\left.+\sum_{s}(W+K)|w_{s}|^{2} 
 -\lambda  \right],
\end{eqnarray}
where $\beta=1/k_{B}T$, and $\omega_{n}=(2n+1)\pi/\beta$ is the fermionic Matsubara frequency.

The most generic mean-field equations do not allow an analytical treatment, although a numerical analysis is straightforward. In order to obtain a deeper physical understanding, we will focus in this paper on the cases in which analytical insight can be gained. This happens when there is only intra-orbital Kondo interactions ($J\neq 0$ and $W=K=0$), which we take up in the remainder of the main text. Furthermore, in presence of particle-hole and time-reversal symmetry~($\lambda=\mu=Q=0$), cases with $W\neq 0$ and/or $K  \neq 0$ also allow for an analytical approach. These are listed in Appendix~\ref{sec:symmetric}. We will not dwell on them because the mean-field equations have a form quite similar to the cases discussed in detail below. 

\begin{figure*}[!t]
\includegraphics[scale=0.9]{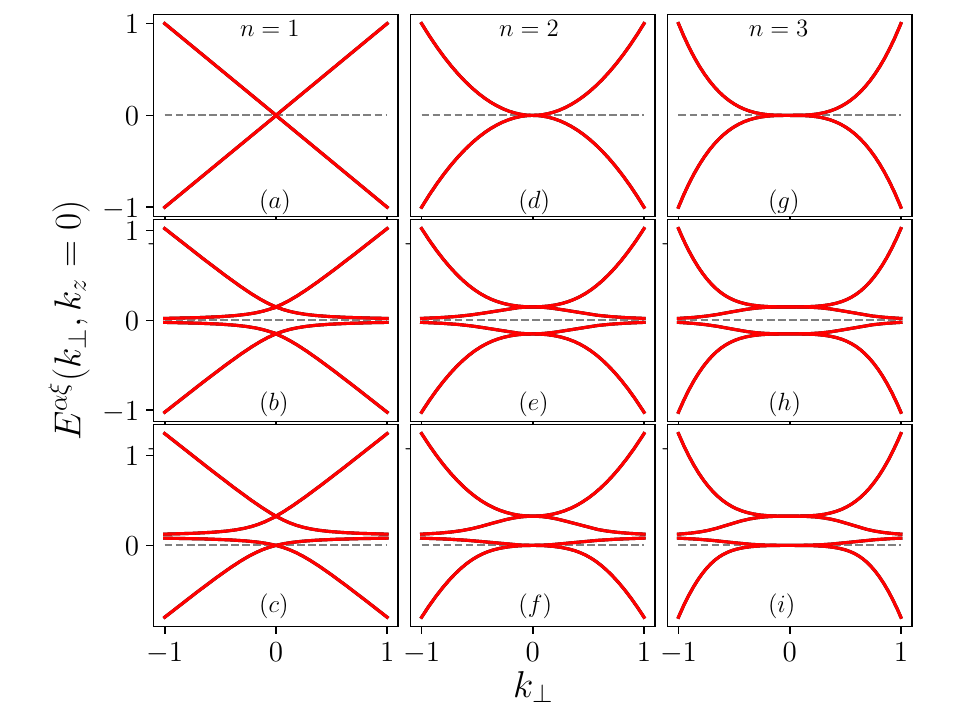}
\caption{Dispersion relations for the multi-Dirac Kondo system (in the $k_{z}=0$ plane) for $n=1,2,3$. Upper panels~[(a),(d),(g))] show the non-hybridized bare dispersions. Middle panels~[((b),(e),(h))] show the multi-Dirac Kondo insulator. Bottom panels~[((c),(f),(i))] show the multi-Dirac Kondo semimetal regime. The parameter values used are $V=0.15\Lambda$, $\lambda=0.1\Lambda$. The horizontal dashed grey line indicates the Fermi energy.
\label{fig1}
}
\end{figure*}

\section{Mean-field theory of the multi-Dirac/Weyl Kondo lattice}\label{sec:multi-Dirac Kondo}

We will explore in detail the 
case in which only the intra-orbital Kondo interaction is present, 
that is $J\neq 0$ and $W=K=0$ in 
the Hamiltonian of Eq.~\eqref{H_Ks}. Using Eqs.~\eqref{H_HS} and~\eqref{F_MF} we obtain a block-diagonal mean-field 
free energy given by
\begin{widetext}
\begin{eqnarray}
 \mathcal{F}_{MF}&=&-\frac{1}{\beta}\sum_{i\omega_{n}}
 \sum_{s}\int\frac{d^{3}\veck}{(2\pi)^{3}}{\rm ln} {\rm Det}\left[-i\omega_{n}\mathbb{1}+
\left(\begin{array}{cc}
                             H_{\veck s}-\mu\mathbb{1} & \mathbb{V}^{s}_{2\times 2} \\
                             {\mathbb{V}}^{*s}_{2\times 2} & \boldsymbol{\lambda}_{2\times 2}
                            \end{array}\right)\right]+\left(\frac{1}{J}\sum_{s}|V_{s}|^{2}-\lambda \right),
\end{eqnarray}
\end{widetext}
where $\mathbb{V}^{s}_{2\times 2}=J\mathbb{v}^{s}_{2\times 2}$. This can be written in terms of the dispersion 
relations as 
\begin{eqnarray}\label{free_energy_J}
 \mathcal{F}_{MF}&=&-\frac{1}{\beta}\sum_{s}
 \int\frac{d^{3}\veck}{(2\pi)^{3}}\sum_{\alpha,\xi}{\rm ln}[1+e^{-\beta E_{s}^{\alpha\xi}(\veck)}] \nonumber \\
 &&+ \left(\frac{1}{J}\sum_{s}|V_{s}|^{2}-\lambda \right),
\end{eqnarray}
where
\begin{equation}\label{bands}
 E_{s}^{\alpha \xi}(\veck)=\frac{1}{2}(\varepsilon_{ks}^{\alpha}-\mu+\lambda)+\frac{\xi}{2}
 \sqrt{(\varepsilon_{ks}^{\alpha}-\mu-\lambda)^{2}+4V_{s}^{2}},
\end{equation}
and
\begin{equation}\label{e_k}
\varepsilon_{ks}^{\alpha}=\alpha\sqrt{A_{n}^{2}k_{\perp}^{2n}+v_{z}^{2}(k_{z}-sQ)^{2}}-sQ_{0},
\end{equation}
with $A_{n}=k_{0}^{(1-n)}v_{\perp}$, $k_{\perp}^{2}=k_{x}^{2}+k_{y}^{2}$, $\alpha=\pm$ and $\xi=\pm$. The mean-field 
parameters $V_{s}=Jv_{s}$ and $\lambda$ are determined by the 
saddle point equations 
\begin{eqnarray}
 \frac{\partial\mathcal{F}_{MF}}{\partial V_{s}}= 
  \frac{\partial\mathcal{F}_{MF}}{\partial \lambda}=0,
\end{eqnarray}
resulting in 
\begin{eqnarray}\label{gap_equation_2}
 &&\int \frac{d^{3}\veck}{(2\pi)^{3}}
 \left[\frac{1}{E_{s}^{++}(\veck)-E_{s}^{+-}(\veck)}[f(E_{s}^{++}(\veck))-f(E_{s}^{+-}(\veck))] \right. \nonumber \\
 &&+
 \left.\frac{1}{{E}_{s}^{-+}(\veck)-{E}_{s}^{--}(\veck)}(\veck)[f({E}_{s}^{-+}(\veck))-f({E}_{s}^{--}(\veck))]\right] \nonumber \\
 &&=-\frac{1}{2J}
\end{eqnarray}
and 
\begin{eqnarray}
 &&\sum_{s,\alpha,\xi}\int \frac{d^{3}\veck}{(2\pi)^{3}}f\left(E_{s}^{\alpha\xi}(\veck)\right)\nonumber \\
 &&\times\left[1-
 \xi\frac{(\varepsilon_{ks}^{\alpha}-\mu-\lambda)}{\sqrt{(\varepsilon_{ks}^{\alpha}-\mu-\lambda)^{2}+4V_{s}^{2}}}\right]=1.
\end{eqnarray}
Above, $f(x)=(1+e^{\beta x})^{-1}$ is the 
Fermi-Dirac distribution.
We will now solve these self-consistent equations at zero temperature for 
different physical situations.

\subsection{The multi-Dirac Kondo lattice (Preserved TRS and IS)}\label{sec:multi-dirac}

\begin{figure}
\includegraphics[scale=0.5]{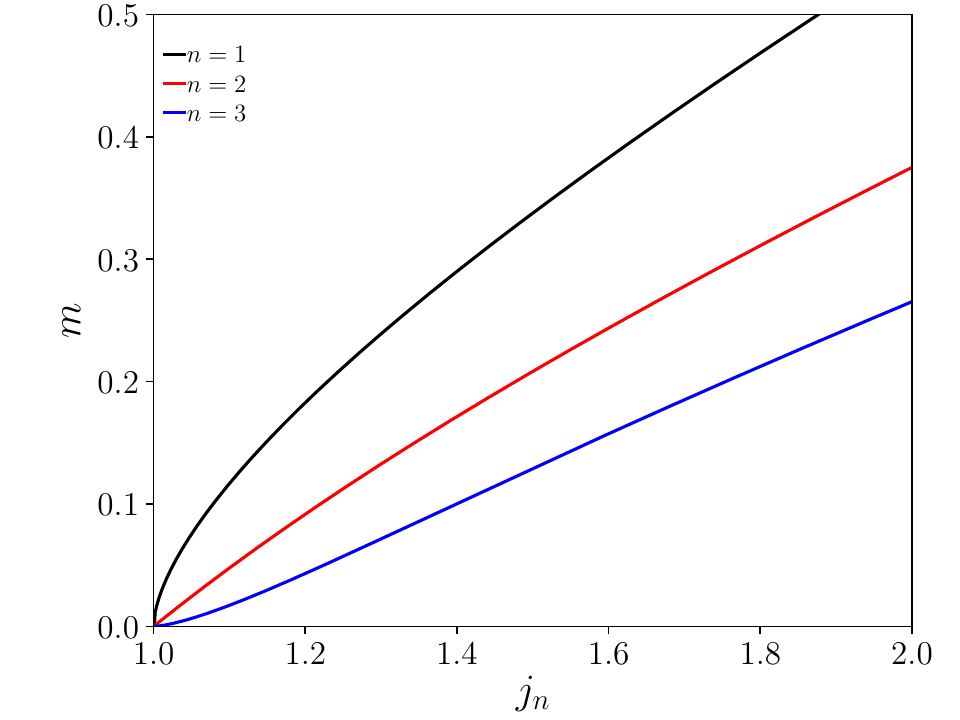}
\caption{ Numerical solution for the multi-Dirac Kondo insulator gap equation~\eqref{gap_dirac_f} for $n=1,2,3$.  
\label{fig2}
}
\end{figure}

When both TRS and IS are present~($Q=Q_{0}=0$) $H_{0}$ in Eq.~\eqref{h_0} describes a 3D multi-Dirac system with gapless Kramers-degenerate Dirac nodes at $\veck_{D}=0$, see Fig.\ref{fig1}[(a),(d),(g)]. First, let us analyze the particle-hole symmetric case, $\mu=\lambda=0$. In the absence magnetic moments, 
the system is a multi-Dirac semimetal. In the 
presence of these moments, however, due to the semimetalic 
nature of the system, a finite mean-field $V_s$ only occurs above a critical value ($J>J_{n}^{c}$), which is characteristic of the pseudo-gap Kondo problem~\cite{PhysRevLett.64.1835}. Therefore, for $J>J_{n}^{c}$
the $f$-electrons hybridize with itinerant multi-Dirac $c$-electrons giving rise to heavy quasiparticles. The Dirac nodes are energetically split, opening an energy gap in the spectra, Fig.\ref{fig1}[(b),(e),(h)]. Thus, 
the system becomes a \textit{multi-Dirac Kondo insulator}. If both TRS an IS are present, we have $V_{s}=V$, and at $T=0$ we obtain the gap equation 
\begin{equation}\label{gap_dirac_f}
 \int_{0}^{1}dx\frac{x^{2/n}}{\sqrt{x^{2}+m^{2}}}=\frac{n}{2j_{n}}.
\end{equation}
Above, $j_{n}=J/J_{n}^{c}$, where $J_{n}^{c}=(2/n)/\rho_{n}(\Lambda)$ is the critical Kondo coupling, $\rho_{n}(\Lambda)$ is the multi-Dirac density of states at the high-energy cutoff $\Lambda$, and $m=2V/\Lambda$ (see Appendix~\ref{sec:kondo_temperature} for details).  
Numerical solutions of the gap equation  for $n=1,2,3$ 
are show in~Fig.~\ref{fig2}. A similar gap 
equation is found in the study of the axionic insulator phase for general multi-Weyl systems~\cite{PhysRevB.95.201102}.

\begin{figure}
\includegraphics[scale=0.5]{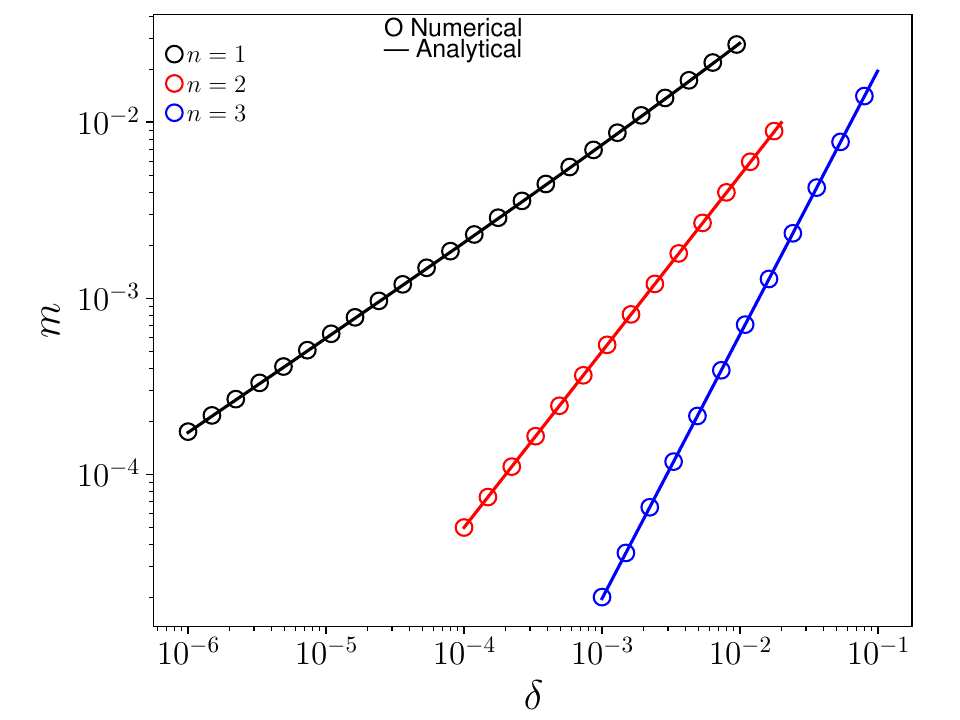}
\caption{ Critical behavior of the multi-Dirac Kondo insulator order parameter $m$ for $n=1,2,3$. Symbols 
show numerical results, while solid lines are the analytical results of Eqs.~\eqref{critical_nt} and \eqref{critica_n1}.  
\label{fig3}
}
\end{figure}

Now, to study in detail the order parameter critical behavior near
the quantum phase transition~(QPT) point $j_{n}\approx 1$ $(J\approx J_{n}^{c})$, we write $j_{n}=1+\delta$, with $\delta=|J-J_{n}^{c}|/J_{n}^{c}\ll 1$. For $n>1$,~(see the details in Appendix~\ref{sec:critical})
\begin{equation}\label{critical_nt}
 m(\delta)=C_{n}\delta^{n/2}, \qquad (n>1)
\end{equation}
with
$C_{n}=\left[\frac{-n\sqrt{\pi}}{\Gamma\left(\frac{1}{2}+\frac{1}{n}\right)\Gamma\left(-\frac{1}{n}\right)}\right]^{n/2}$. Specifically,  $C_{2}=1$, $C_{3}\approx 1.249$. In the single-Dirac case~($n=1$) the critical behavior is modified by
 \begin{eqnarray}\label{critica_n1}
m(\delta)=f(\delta)\delta^{1/2},\qquad (n=1),  
 \end{eqnarray}
where $f(\delta)$ carries logarithmic corrections, calculated in Appendix~\ref{sec:critical}. The 
comparison between numerical solutions and the analytic 
expressions \eqref{critical_nt} and \eqref{critica_n1} 
is shown in Fig.~\ref{fig3}. The mean-field quantum critical exponent is thus $\nu=n/2$, in agreement with early work~\cite{PhysRevLett.64.1835}.
\begin{figure}
\includegraphics[scale=0.57]{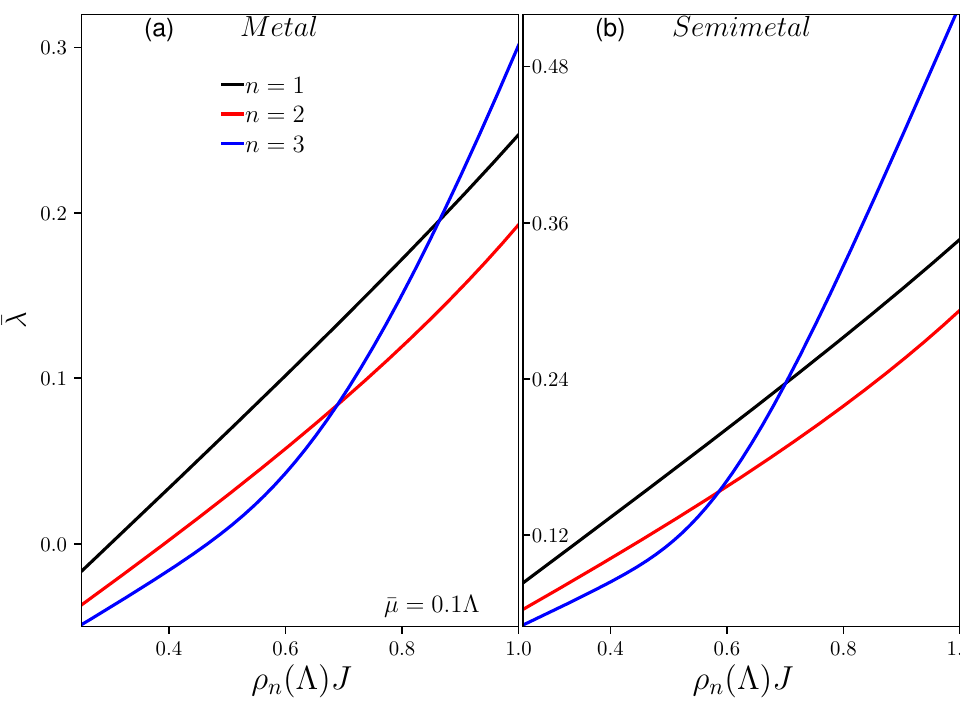}
\caption{Numerical solution of the self-consistent mean-field parameter $\bar{\lambda}$
for $n=1,2,3$ for (a) the multi-Dirac Kondo metal 
and (b) 
the multi-Dirac Kondo semimetal.
\label{fig4}
}
\end{figure}
\begin{figure}
\includegraphics[scale=0.55]{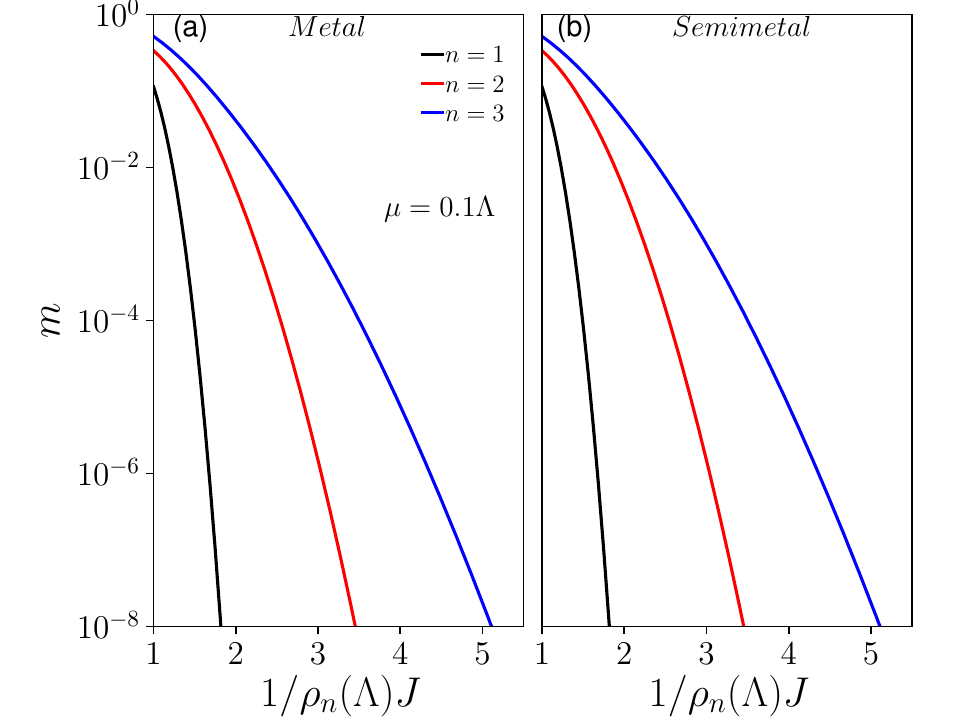}
\caption{Semi-log plot of the numerical solution of the mean-field parameter $m$ for the multi-Dirac Kondo metal and semimetal for $n=1,2,3$.
\label{fig5}
}
\end{figure}

%
\begin{figure*}
\includegraphics[scale=1.2]{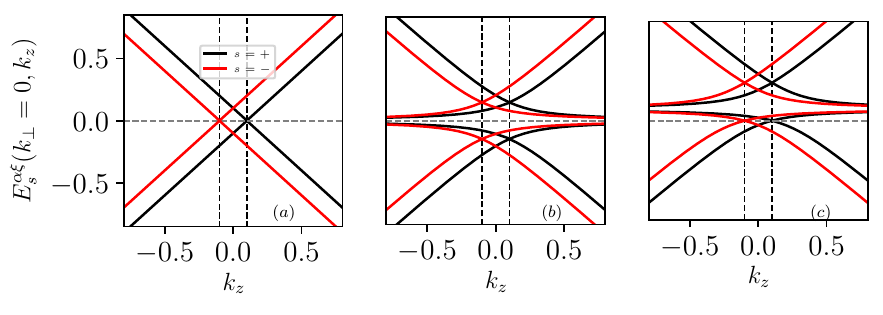}
\caption{Dispersion relations for the broken-TRS multi-Weyl Kondo system (in the $k_{\perp}=0$ plane), for $n=1$. (a) The non-hybridized case~($J=0$). (b) The broken-TRS multi-Weyl Kondo insulator~($\lambda=\mu=0$). (c) The broken-TRS multi-Weyl Kondo semimetal~($\mu=-V^{2}/\lambda$).  The parameters used are 
$Q=0.1\Lambda$, $V=0.15\Lambda$, $\lambda=0.1\Lambda$. The horizontal dashed grey line indicates the Fermi energy.
The vertical dashed black lines indicate the Weyl node $k$-points $k_{W}^{\pm}=(0,0\pm Q)$. 
\label{fig6}
}
\end{figure*}

Away from particle-hole symmetry~($\mu\neq 0$), the density of states of the unhybridized system~($J=0$) is always finite and  the Kondo effect occurs for any finite Kondo coupling. 
Therefore, we obtain a \textit{multi-Dirac Kondo metal}. The 
self-consistent equations at $T=0$ are given by 
\begin{subequations}
\label{gap_eq_11}
\begin{eqnarray}
 \int_{0}^{1}dx\frac{x^{(2/n)+1}}{\sqrt{(x-\alpha)^{2}+m^{2}}}&=&\frac{1}{\rho_{n}(\Lambda)\Lambda}\left[1-\frac{\alpha}{J/\Lambda}\right], \\
  \int_{0}^{1}dx\frac{x^{2/n}}{\sqrt{(x-\alpha)^{2}+m^{2}}}&=&\frac{1}{\rho_{n}(\Lambda)J}.
 \end{eqnarray}
 \end{subequations}
where $\alpha=(\mu+\lambda)/\Lambda$~(see Appendix~\ref{sec:kondo_temperature}). 

A particular case of special interest occurs when $\mu=-V^{2}/\lambda$~($\lambda>0$). In this case, the  lower hybridized
Dirac node is pinned to the Fermi energy: 
$E^{+-}_s(\mathbf{k}_{D})=E^{--}_s(\mathbf{k}_{D})=0$. 
This particular choice of chemical potential realizes 
a \textit{multi-Dirac Kondo semimetal}, see Fig.\ref{fig1}[(c),(f),(i)].
This is reminiscent of the choice of Ref.~\cite{doi:10.1073/pnas.1715851115} for the single Kondo-Weyl semimetal.  The self-consistency equations 
of the multi-Dirac Kondo semimetal are given by~\eqref{gap_eq_11}, but with $\alpha=\alpha(m,\bar{\lambda})=-m^{2}/4\bar{\lambda}+\bar{\lambda}$, where $m=2V/\Lambda$, $\bar{\lambda}=\lambda/\Lambda$. The numerical solution of the self-consistent equations both for the multi-Dirac Kondo metal and semimetal are shown in~Figs.~\ref{fig4} and \ref{fig5}. Their behavior is quite similar.



\subsection{The multi-Weyl Kondo lattice (Broken TRS or IS)}
\noindent

\subsubsection{Broken-TRS multi-Weyl Kondo lattice}
\label{brokenTRS}
\noindent

When TRS is broken, and IS is preserved~($Q\neq0$, $Q_{0}=0$)
the Kramers-degenerate Dirac nodes of the unhybridized bands split in momentum space into two multi-Weyl nodes with opposite chiralities at the same energy and distinct $k$-points $\mathbf{k}_{W}^{\pm}=(0,0,\mp Q)$~\cite{PhysRevB.85.165110,PhysRevLett.111.027201}, see Fig.~\ref{fig6}(a). If the energy of those Weyl nodes coincides with the Fermi energy we have a nodal semimetal before coupling to the local moments. For strong enough interactions with the lattice of local moments and at the particle-hole symmetric point~($\mu=\lambda=0$), the system becomes a \textit{broken-TRS multi-Weyl Kondo insulator}, because the hybridization opens a gap at the Weyl nodes, see Figs.\ref{fig6}(b). At first, it would appear that the hybridized bands would define two distinct order parameters, $V_{\pm}$. After integration of 
the self-consistent equations, however, as in the multi-Dirac Kondo case, $V_{s}=V$~(see Appendix~\ref{sec:kondo_temperature}). 
Thus, the mean-field equations of this broken-TRS multi-Weyl Kondo lattice have a form similar to those of the previously  analyzed multi-Dirac Kondo lattice. 
In particular, if we now, as we did in the multi-Dirac Kondo lattice case, set the chemical potential at $\mu=-V^{2}/\lambda$, we obtain a \textit{broken-TRS multi-Weyl Kondo semimetal}, see Fig.~\ref{fig6}(c).


\subsubsection{Broken-IS multi-Weyl Kondo lattice}
\noindent

\begin{figure}
\includegraphics[scale=0.6]{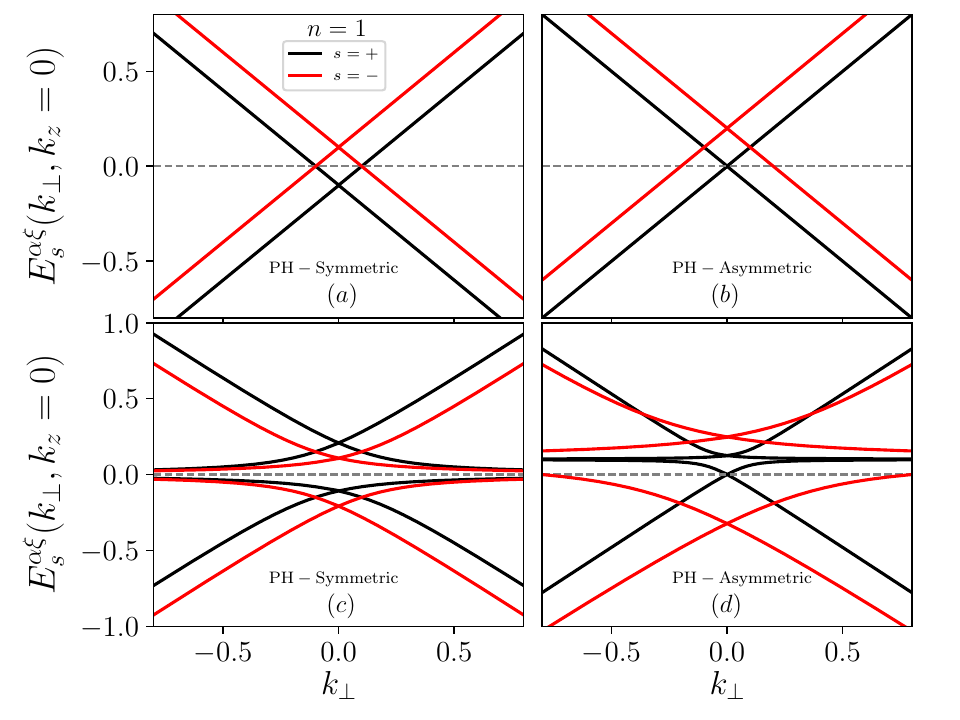}
\caption{Dispersion relations for 
the broken-IS ($Q_{0}=-0.1\Lambda$) single-Weyl Kondo system~(in the $k_{z}=0$ plane). The 
non-hybridized limit~($J=0$) is shown in the upper panels (a) and (b), with ($\mu=0$) and without ($\mu=-Q_{0}$) particle-hole symmetry, respectively. The interacting regimes are shown in the bottom panels (c) and (d), with $\mu=0$ and $\mu=-V_{+}^{2}/\lambda-Q_{0}$, respectively. The parameters used in the interacting particle-hole symmetric case are $V_{+}=V_{-}=0.15\Lambda, \lambda=0$, whereas in the asymmetric case we have $V_{+}=0.05\Lambda$, $V_{-}=0.25\Lambda,\lambda=0.1\Lambda$. The horizontal dashed grey line shows the Fermi energy. 
\label{fig7}
}
\end{figure}
\begin{figure}
\includegraphics[scale=0.6]{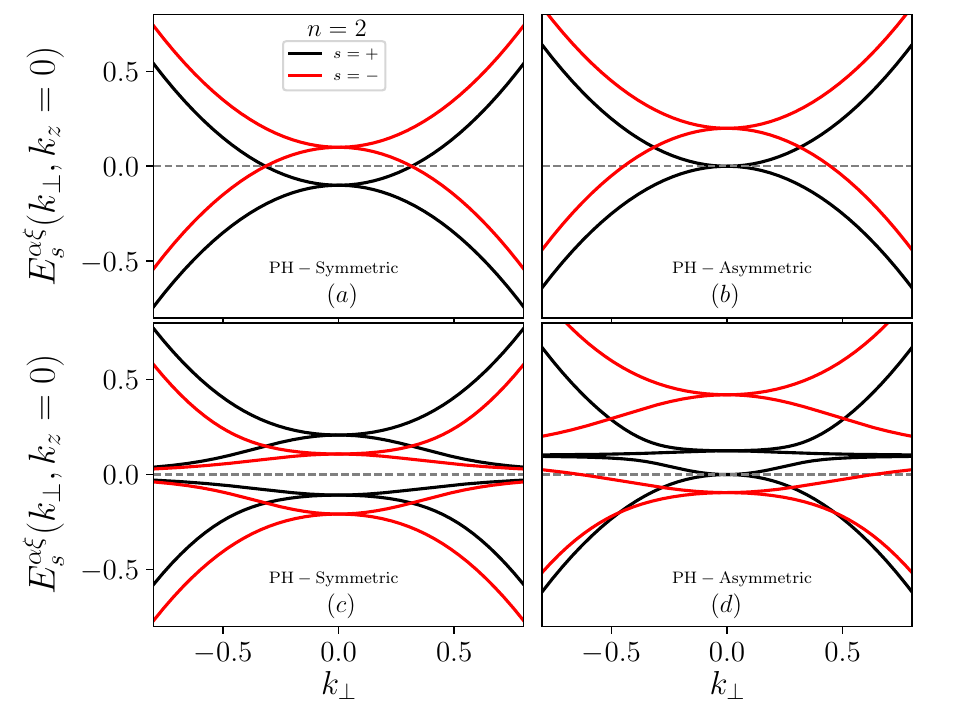}
\caption{Dispersion relations for 
the broken-IS double-Weyl Kondo system~(in the $k_{z}=0$ plane). The description and parameters used are the same as those of Fig.~\ref{fig7}.
\label{fig8}
}
\end{figure}
\begin{figure}[t!]
\includegraphics[scale=0.5]{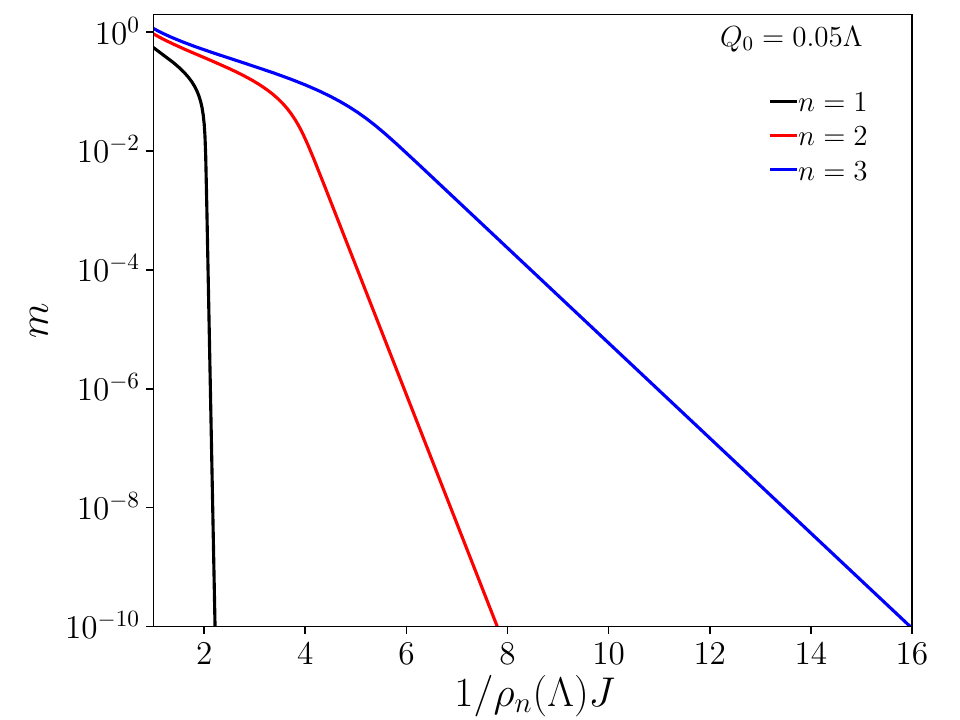}
\caption{Semi-log plot of the order parameter $m$ for the broken-IS multi-Weyl Kondo insulator ($Q_{0}=0.05\Lambda$) for $n=1,2,3$.
\label{fig9}
}
\end{figure}

Finally, we consider the case where there is TRS but IS is broken~($Q=0,Q_{0}\neq 0$). In this case, the 
multi-Weyl nodes occur at the same $k$-point $\mathbf{k}_{W}=\mathbf{0}$, but with different energies $\varepsilon_{\pm W}=\mp Q_{0}$~\cite{PhysRevB.85.165110,PhysRevLett.111.027201}. Thus, IS breaking destroys the nodal semimetal phase generically producing a phase  with electron and hole Fermi surfaces~\cite{PhysRevB.85.165110}. The dispersion relations for the non-hybridized broken-IS single and 
double-Weyl metals are shown in the upper panels of Fig.~\ref{fig7} ($n=1$) and \ref{fig8} ($n=2$).  In Figs.~\ref{fig7}(a) and \ref{fig8}(a), the 
particle-hole symmetric case~($\mu=0$)  is shown, whereas in
Figs.~\ref{fig7}(b) and \ref{fig8}(b), particle-hole symmetry is broken by setting $\mu=-Q_{0}$, which pins the Weyl node with positive chirality to the Fermi energy. 


We first analyze the effect of coupling the lattice of local moments to the broken-IS multi-Weyl Kondo lattice at  particle-hole symmetry ($\mu=\lambda=0$).
In contrast to the multi-Dirac and broken-TRS multi-Weyl cases, now the Fermi surface is always finite and the local moments Kondo-bind to the itinerant $c$-electrons for any finite Kondo coupling: thus, the critical value $J_c=0$.  Because a  gap opens at the chemical potential, we will call this a \textit{broken-IS multi-Weyl Kondo insulator}, as shown in  Figs.~\ref{fig7}(c) and \ref{fig8}(c).
Moreover,
 $V_{+}=V_{-}=V$, and, at $T=0$, $f(E_{\veck s}^{++})=f(E_{\veck s}^{-+})=0$, and $f(E_{\veck s}^{+-})=f(E_{\veck s}^{--})=1$, so the mean-field equation becomes
\begin{eqnarray}\label{gap_eq_insu}
&& \sum_{s=\pm}\int_{0}^{1}dx\frac{x^{2/n}}{\sqrt{(x-s\bar{Q}_{0})^{2}+m^{2}}}=\frac{2}{\rho_{n}(\Lambda)J}.
\end{eqnarray}
Here, $m=2V/\Lambda$, $\bar{Q}_{0}=Q_{0}/\Lambda$.
Numerical solutions for the order parameter $m$ for $n=1,2,3$ are shown in Fig.~\ref{fig9}, where the usual Kondo exponential dependence can be discerned. 
\begin{figure*}[!t]
\includegraphics[scale=1.0]{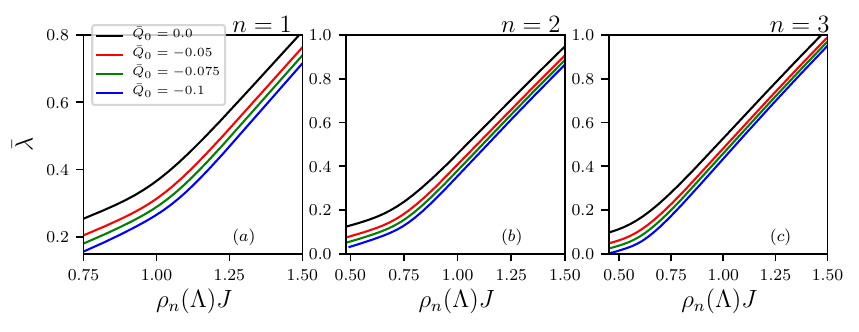}
\caption{Self-consistent parameter $\bar{\lambda}$ for the broken-IS multi-Weyl Kondo lattice for $n=1,2,3$ for several values of $\bar{Q}_{0}$. 
\label{fig10}
}
\end{figure*}
\begin{figure}[!t]
\includegraphics[scale=0.55]{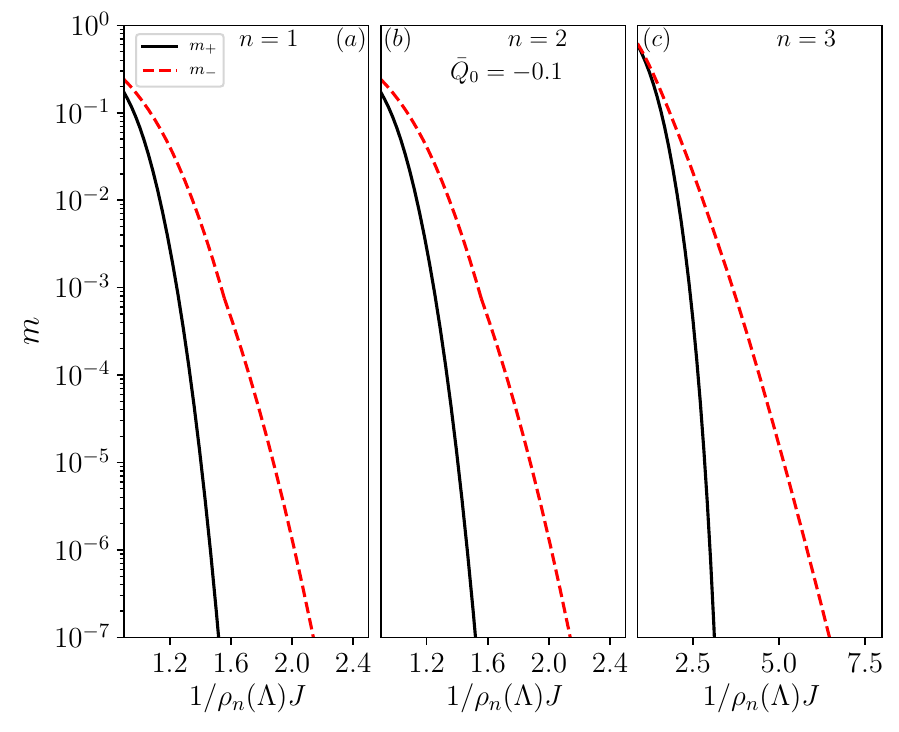}
\caption{Semi-log plot of Kondo parameters $m_{+}$ and $m_{-}$ for the broken-IS multi-Weyl Kondo lattice for $n=1,2,3$ for $\bar{Q}_{0}=-0.1$. 
\label{fig11}
}
\end{figure}
\begin{figure}[!t]
\includegraphics[scale=0.5]{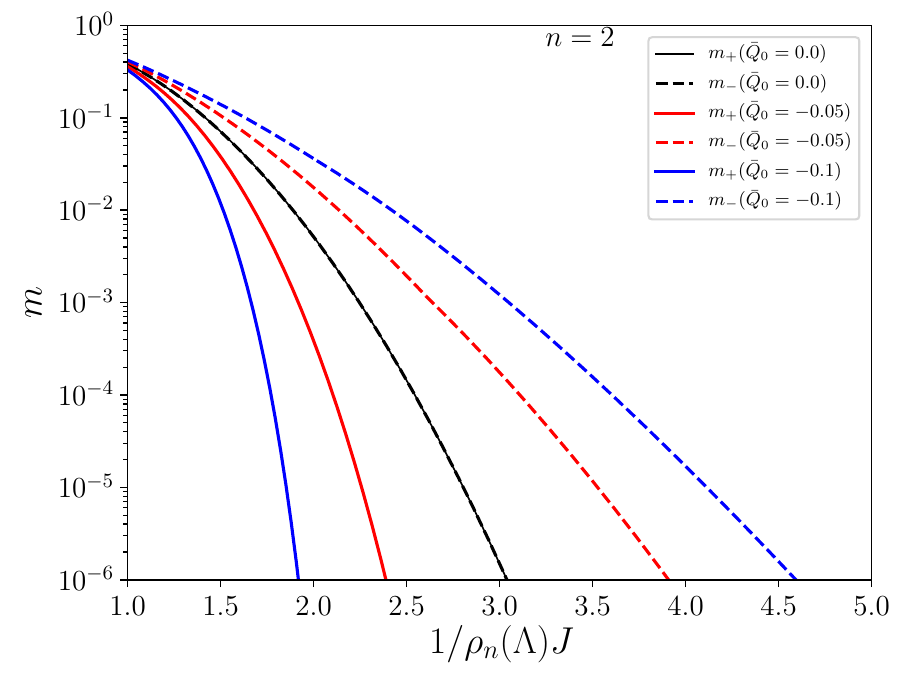}
\caption{Semi-log plot of Kondo parameters $m_{+}$ and $m_{-}$ for the broken-IS multi-Weyl Kondo lattice for $n=2$ for several values of $\bar{Q}_{0}$. 
\label{fig12}
}
\end{figure}

In the limit $\rho_{n}(\Lambda)J\ll 1$ and $m\ll \bar{Q}_{0}$, it is possible to extract analytically the exponential dependence on $J$ for $n=1,2$. We find
\begin{subequations}
\label{m_KI}
\begin{eqnarray}
m&=&f_{1}^{KI}(Q_{0}/\Lambda) {\exp}[-1/(2\rho^{W}_{1}(0)J)] \ (n=1), \\ 
m&=&f_{2}^{KI}(Q_{0}/\Lambda){\exp}[-1/(2\rho_{2}^{W}(0)J] \ (n=2),
\end{eqnarray}
\end{subequations}
where
\begin{subequations}
\begin{eqnarray}
f_{1}^{KI}(x)&=&\sqrt{\frac{x(1-x)}{2}}{\rm exp}[(1-3x^{2})/2x^{2}], \\
f_{2}^{KI}(x)&=&\sqrt{\frac{x(1-2x)}{2}}{\rm exp}[(2-2x)/2x],
\end{eqnarray}
\end{subequations}
and $\rho_{n}^{W}(0)$ is the multi-Weyl 
density of states at the Fermi energy~(see Appendix~\ref{sec:weyl_dos} for details).
As expected, since IS breaking drives the critical coupling $J_c$ to zero, the order parameter is non-analytic in both $J$ and $Q_0$. 
The explicit exponential dependence of $m$  on $Q_{0}$ becomes clear when we note that $\rho_{n}^{W}(0)\propto Q_{0}^{2/n}$~(see Eq.\eqref{dos_ss} and Eq.\eqref{multi-dos} of Appendix~\ref{sec:weyl_dos}).
Moreover, it is reminiscent of the Kondo temperature obtained previously by one of us using the numerical renormalization group in the study of multi-Weyl systems in the presence of a single quantum impurity~\cite{PhysRevB.103.045137}.
We have checked that these analytical expressions perfectly match the numerical results in their validity region.


%
 
We now consider the particle-hole asymmetric case. In this case, $V_{+}\neq V_{-}$. Inspired by the choice of Ref.~\cite{doi:10.1073/pnas.1715851115}, we set the chemical potential to $\mu=-(V^{2}_{+}/\lambda)-Q_{0}$, which pins the unhybridized multi-Weyl node with positive chirality to the Fermi energy, i.e. $E_{+}^{+-}(\mathbf{k}_{W})=E_{+}^{--}(\mathbf{k}_{W})=0$, see Figs.~\ref{fig7}(d) and \ref{fig8}(d). Unlike the broken-TRS multi-Weyl Kondo lattice, in broken-IS Weyl systems with both Weyl nodes at the same $k$-point, as we have here, a nodal semimetallic phase is not possible. This is because, at the Fermi energy, although the density of states associated with a particular Weyl chirality might go to zero, the density of states associated with the opposite chirality is always non-vanishing. Moreover, as we will see, this difference will lead to a huge difference between the mean-field order parameters associated with each chirality.

The self-consistency equations in the broken-IS particle-hole asymmetric Weyl-Kondo lattice setting $\mu=-(V_{+}^{2}/\lambda)-Q_{0}$ are~(see Appendix~\ref{sec:kondo_temperature} for details)
\begin{subequations}
\label{coupled_MF_2}
\begin{eqnarray}
 &&\int_{0}^{1}dx\frac{x^{(2/n)+1}}{\sqrt{\left[x+\alpha(m_{+},\bar{\lambda})\right]^{2}+m_{+}^{2}}} \nonumber \\
 &&+\int_{0}^{1}dx\frac{x^{(2/n)+1}}{\sqrt{\left[x+\alpha(m_{+},\bar{\lambda})+2\bar{Q}_{0}\right]^{2}+m_{-}^{2}}} \nonumber \\
 &&=\frac{2}{\rho_{n}(\Lambda)\Lambda}\left[1-\frac{\alpha(m_{+},\bar{\lambda})+\bar{Q}_{0}}{J/\Lambda}\right], \\
 &&\int_{0}^{1}dx\frac{x^{2/n}}{\sqrt{\left[x+\alpha(m_{+},\bar{\lambda})\right]^{2}+m_{+}^{2}}}=\frac{1}{\rho_{n}(\Lambda)J}, \\
 &&\int_{0}^{1}dx\frac{x^{2/n}}{\sqrt{\left[x+\alpha(m_{+},\bar{\lambda})+2\bar{Q}_{0}\right]^{2}+m_{-}^{2}}}=\frac{1}{\rho_{n}(\Lambda)J}. \nonumber \\
\end{eqnarray}
\end{subequations}
Above, $m_{\pm}=2V_{\pm}/\Lambda$, $\bar{Q}_{0}=Q_{0}/\Lambda$, and $\alpha(m_{+},\bar{\lambda})=-(m_{+}^{2}/4\bar{\lambda})+\bar{\lambda}$.
Note that, for $Q_{0}=0$, IS is restored and Eqs.~\eqref{coupled_MF_2} reduce to Eqs.~\eqref{gap_eq_11} for the multi-Dirac Kondo case, as expected. 
Besides, for $Q_{0}\neq 0$, like in the particle-hole symmetric case, the total density  of states at the Fermi level is finite and the critical coupling constant is zero. 
The numerical solutions of Eqs.~\eqref{coupled_MF_2} are plotted in Figs.~\ref{fig10} to \ref{fig12}.
Like in the multi-Dirac and broken-TRS multi-Weyl lattice, the self-consistent parameter $\bar{\lambda}$ does not present any exponential behavior, nor does it change significantly with the increase of the IS-breaking parameter $\bar{Q}_{0}$. 
In contrast, from Fig.~\ref{fig11} we can see 
that the order parameters $m_{+}$ and $m_{-}$ both behave exponentially with a marked hierarchy: $m_+\ll m_-$. In Fig.~\ref{fig12}, we show the Kondo parameters for the broken-IS multi-Weyl Kondo lattice for several values of $\bar{Q}_{0}$ for $n=2$. Clearly, although they are equal when there IS ($\bar{Q}_{0}=0$), they differ significantly as $\bar{Q}_{0}=0$ is increased, even though they both decrease exponentially for decreasing $\rho_{n}(\Lambda)J$. 
This is a reflection of the semimetallic character of the positive chirality electronic band, contrasting with the fully metallic nature of the negative chirality sector.

\section{Discussion}
\label{sec:discussion}

The mean-field description of heavy fermion metals~\cite{G.R.Stewart1984} and Kondo insulators~\cite{riseboroughreview} has played a pivotal role in our understanding of these large families  of compounds. Complications due to magnetic ordering~\cite{Doniach1977}, in particular quantum critical behavior~\cite{Gegenwart2008}, and heavy fermion superconductivity~\cite{Heffner1996} are beyond the scope of the theory. However, on the paramagnetic side of the Doniach phase diagram~\cite{Doniach1977}, considerable insight has been gained, from the origin of the heavy effective masses to the nature of the renormalized Kondo insulating gap. This work offers an analogous phenomenology for multi-Dirac and multi-Weyl Kondo systems. 

We have seen that, at particle-hole symmetry, the bare multi-Dirac and TRS-broken multi-Weyl densities of states have  the power law form $\rho(\omega)\propto \omega^{r}$, with $r=2/n$~(see Eq.\eqref{multi-dirac-dos} in Appendix~\ref{sec:weyl_dos}). Then, in the presence of Kondo spins, these systems become archetypes of pseudogap Kondo systems. The pseudogap problem for a single Kondo impurity was originally proposed by Withoff and Fradkin (WF)~\cite{PhysRevLett.64.1835}. That and subsequent works focused on a large-N approach similar to the one we used here~\cite{CassanelloFradkin1,CassanelloFradkin2,Polkovnikov}.
The problem has also been extensively analyzed using the perturbative renormalization group~(PRG)~\cite{PhysRevB.65.014511,PhysRevB.69.174421,PhysRevB.70.094502,PhysRevB.70.214427}  and the  numerical renormalization group (NRG)~\cite{Chen1995,PhysRevB.54.11936,Bulla1997,PhysRevB.57.14254,PhysRevLett.89.076403}. It has been thoroughly reviewed in~\cite{vojtaimpurityreview}. 
Although we are dealing here with a lattice of spins instead of a single impurity, the local nature of the large-$N$ inspired approach should lead to similar results. In particular, the critical behavior close to $J_c$ should be similar. In this respect, subtle complications arise due to the absence of exact particle-hole symmetry \emph{ at all scales} in any realistic description~\cite{CassanelloFradkin1,CassanelloFradkin2,Polkovnikov,vojtaimpurityreview}. We have skirted these issues by working within a simplified particle-hole symmetric description close to the Dirac/Weyl nodes that ignores higher-energy  deviations from that symmetry. This simplification leads to the critical behavior we found for the multi-Dirac Kondo in Section~\ref{sec:multi-dirac} and the TRS-broken multi-Weyl case in Sectionn~\ref{brokenTRS}, namely, that the characteristic energy scale $m\propto |J-J_{c}|^{\nu}$, with $\nu=1/r$. We expect that a more microscopic description of the Dirac/Weyl bands would change the detailed critical behavior but not the presence of a critical coupling constant in the cases considered. 

If the coupling constant exceeds the critical value, then the usual opening of a Kondo insulating gap ensues. In contrast to usual Kondo insulators~\cite{riseboroughreview}, though, the renormalized energy gap is not an exponential function of $J$. This may have detectable experimental consequences, e.g., in pressure studies, which can tune the microscopic value of $J$. Away from particle-hole symmetry, the generic behavior is that of a heavy fermion metal, with the usual phenomenology. 

A particular fine-tuned case of possible interest, however, occurs when the chemical potential of the multi-Dirac or TRS-broken multi-Weyl Kondo lattice is at or close to the Dirac or Weyl nodes, see Figs.~\ref{fig1}[(c),(f),(i)] and \ref{fig6}(c). Then, the hybridized multi-Dirac/Weyl nodes will have renormalized velocities, a consequence of the composite nature of fermions in the Kondo condensate. As a result, physical quantities such as the specific heat will be deeply affected. In general, it will depend on temperature as a higher power law than in a metal
\begin{equation}
 \frac{C_{v}(T)}{T}=\gamma_{n}T^{2/n},
\end{equation}
where $\gamma_n$ will reflect the 
nodal velocity renormalization~(see Appendix~\ref{sec:specific}). 
Thus, specific heat measurements could play an important role in the search for signatures of strongly correlated topological materials~\cite{Dzsaber2021,Chen2022}.



In sharp contrast, for broken-IS multi-Weyl systems, the bare density of states is always non-zero at the Fermi level, see Fig.\ref{fig7} and Fig.\ref{fig8}, and Kondo condensation will occur for any value of $J$. At particle-hole symmetry, a broken-IS multi-Weyl Kondo insulator ensues, see Fig.\ref{fig7}(c) and Fig.\ref{fig8}(c). The renormalized Kondo gap has the usual exponential dependence on $J$, as can be seen in~Fig.~\ref{fig9} and from Eq.~\eqref{m_KI}. Note, however, that the IS-breaking momentum $Q_0$ also appears in the argument of the exponential, which might offer an opportunity for experimental detection if $Q_0$ can be externally tuned.

The heavy fermion metal phenomenology applies away from particle-hole symmetry, but now there are different mean-field order parameters $m_+\neq m_-$, each one responsible for renormalizing a different chirality sector. Even for modest IS-breaking, these two can be parametrically widely separated, as seen in Figs.~\ref{fig11} and \ref{fig12}.
These widely different energy scales are expected to be reflected in the temperature dependence, the smallest $m_s$ crossing over to zero at a much lower temperature than the other. Evidently, these should give rise to detectable consequences in thermodynamic and transport properties. 

If the chemical potential happens to fall at or close to one of Weyl nodes there will be both metallic and semimetallic bands, as shown in Figs.~\ref{fig7}(d) and \ref{fig8}(d). The specific heat will be given by $C_{V}/T=\gamma_{0} +\gamma_{n}T^{2/n}$, the first and second terms coming from the metallic and semimetallic bands, respectively~(see Appendix~\ref{sec:specific}).
Whether the presence of these two terms can be distinguished from other contributions, such as phonons, remains to be seen.

One final remark is worth making. Whenever any of the order parameters $m_s$ has the familiar exponential dependence on $1/J$, we expect its value to be on the order of the ratio between the Kondo temperature and the band width, $T_K/D$, in conventional heavy-fermion materials, namely, $m_s \sim 10^{-3}-10^{-2}$. The inverse of this ratio will determine the renormalizations of the corresponding quasiparticle masses  or Kondo insulating gap, whichever applies. In the semimetallic phases we analyzed here, however, the Dirac/Weyl masses are only weakly renormalized, as can be easily shown. This is because we chose a model with Dirac/Weyl points \emph{at the zone center}. Had we chosen them closer to the zone boundaries, they would also have been strongly renormalized on a $1/m_s$ scale (see, e.g., references \cite{PhysRevLett.118.246601,doi:10.1073/pnas.1715851115,PhysRevB.101.075138,Dzsaber2021,doi:10.1073/pnas.1715851115,PhysRevB.104.085151}).

\section{Concluding Remarks}
\label{sec:conclusions}
In this manuscript we have studied multi-Dirac/Weyl systems in the presence of a lattice of local moments using the multi-Dirac/Weyl Kondo lattice model. We have done so based on the large-$N$ inspired mean-field approach, adequate to the paramagnetic side of the Doniach phase diagram. At particle-hole symmetry and in the multi-Dirac and broken-TRS Weyl cases, there is a non-zero coupling constant $J_c$, below which no Kondo compensation occurs. For $J>J_c$, however, there is Kondo quenching and Kondo insulating behavior, but with a non-exponential characteristic energy scale. Away from particle-hole symmetry $J_c=0$, and either heavy fermion metallic or semimetallic behavior can occur, depending on the position of the chemical potential. For the broken-IS multi-Weyl Kondo lattice $J_c=0$ always. Depending on the conduction electron filling, we can have a Kondo insulator or a heavy fermion metal. Different quasi-particle renormalizations are expected for different chirality sectors.  

Our results have experimental consequences that could in principle be sought in some compounds. In particular, we have found a complex phenomenology for the mass and velocity renormalizations of the quasi-particles, which we have described in detail, that could be directly investigated. More detailed studies of transport properties and the effects of disorder are promising future directions to pursue. 
Beyond mean-field theory, there remain important questions we have not touched, such as magnetic ordering, quantum criticality and superconductivity. We hope our results will serve as an initial guide in those directions.

\acknowledgements

J. F. S acknowledges the support of the postdoctoral fellowship from FAPESP No. 2021/04078-0. E. M. acknowledges the support of CNPq through Grant No. 309584/2021-3 and Capes through Grant No. 0899/2018.

\appendix
\section{The Multi-Weyl density of states}
\label{sec:weyl_dos}

The multi-Weyl density of states can be computed using the multi-Weyl matrix Green's function $G_{\veck}(\omega)=[(\omega +\mu)\mathbb{1}-\mathcal{H}_{\veck}]^{-1}$, where the 
Hamiltonian $\mathcal{H}_{\veck}$ is given by Eq.~\eqref{h_kk}. The spin-diagonal part of the Green's function is given by 
\begin{eqnarray}
 G_{\veck \s s}^{W}(\omega)&=&\frac{\omega+\mu+sQ_{0}}{[\omega+\mu+sQ_{0}]^{2}-\epsilon_{ks}^{2}}
 \nonumber \\
 &+&s\sigma
 \frac{v_{z}(k_{z}-sQ)}{[\omega+\mu+sQ_{0}]^{2}-\epsilon_{ks}^{2}}.
\end{eqnarray}
Above, $\s=\pm$, $s=\pm$, $\epsilon_{ks}=\sqrt{A^{2}_{n}k_{\perp}^{2n}+v_{z}^{2}(k_{z}-sQ)^{2}}$, with $k_{\perp}^{2}=k_{x}^{2}+k_{y}^{2}$, and $A_{n}=k_{0}^{(1-n)}v_{\perp}$. The spectral function is
\begin{eqnarray}
 &&A_{\veck \s s}^{W}(\omega)=\frac{1}{\pi}{\rm Im}[G_{\veck \s s}^{W}(\omega-i\delta)] \nonumber \\
 &&=|\omega+\mu+sQ_{0}|\delta[(\omega+\mu+sQ_{0})^{2}-\epsilon_{ks}^{2}] \nonumber \\
 &&+s\s \frac{v_{z}(k_{z}-sQ)}{2\varepsilon_{ks}}\left[\delta(\omega+\mu+sQ_{0}-\varepsilon_{ks})\right.\nonumber \\
&&\left. -\delta(\omega+\mu+sQ_{0}+\varepsilon_{ks})\right]. \nonumber \\
\end{eqnarray}
The density of states is
\begin{eqnarray}\label{dos_component}
  \rho_{\sigma s}^{W}(\omega)&=&\int\frac{d^{3}\veck}{(2\pi)^{3}} A_{\veck \s s}^{W}(\omega) \nonumber \\
  &=&\frac{1}{(2\pi)^{2}}\int_{-\Lambda^{\prime}}^{\Lambda^{\prime}}dk_{z}\int_{0}^{\Lambda^{\prime}}dk_{\perp}k_{\perp}A_{\veck \s s}^{W}(\omega), \nonumber \\
\end{eqnarray}
where $\Lambda^{\prime}$ is a high-momentum cutoff. The density of states is independent of the TRS-breaking parameter $Q$. To see that, we shift $k_{z}-sQ\rightarrow k_{z}$, which leads to 
\begin{eqnarray*}
  \rho_{\sigma s}^{W}(\omega)
  &=&\frac{1}{(2\pi)^{2}}\int_{-\Lambda^{\prime}-sQ}^{\Lambda^{\prime}-sQ}dk_{z}\int_{0}^{\Lambda^{\prime}}dk_{\perp}k_{\perp}A_{\veck \s s}^{W}(\omega), \nonumber \\
\end{eqnarray*}
from which we can drop $Q$ in the limit of moderate TR breaking $\Lambda^{\prime} \gg Q$. We now make the change of variables
\begin{equation}\label{changes}
A_{n}k_{\perp}^{n}=\rho \cos \theta, \quad k_{z}v_{z}=\rho \sin\theta,
\end{equation}
and use a different regularization at high energies (which does not affect the low energy physics), with $0 \leq \rho\leq \Lambda, 
-\pi/2\leq \theta \leq \pi/2$ and
$$dk_{z}dk_{\perp}=\frac{\rho^{1/n}}{nA_{n}^{1/n}v_{z}}\cos^{({\frac{1}{n}-1})}(\theta)d\rho d\theta.$$
The integrals in Eq.~\eqref{dos_component} can now be easily computed as
\begin{eqnarray}\label{int1}
&& \int\frac{d^{3}\veck}{(2\pi)^{3}}\delta[(\omega+\mu+sQ_{0})^{2}-\varepsilon_{ks}^{2}] \nonumber \\
&&=\frac{1}{(2\pi)^{2}nA_{n}^{2/n}v_{z}}\frac{\sqrt{\pi}\Gamma(1/n)}{\Gamma\left(\frac{1}{2}+\frac{1}{n}\right)}\frac{|\omega+\mu+sQ_{0}|^{(2/n)-1}}{2}, \nonumber \\
\end{eqnarray}
where $\Gamma(z)$ is the Gamma function,
and 
\begin{eqnarray}\label{int2}
 &&\int\frac{d^{3}\veck}{(2\pi)^{3}}\frac{(k_{z}-sQ)}{2\varepsilon_{ks}}\left[\delta(\omega+\mu+sQ_{0}-\varepsilon_{ks}) \right.\nonumber \\
 &&\left.-\delta(\omega+\mu+sQ_{0}-\varepsilon_{ks})\right]=0.
\end{eqnarray}
The final expression for the density of states per spin and chirality is
\begin{equation}\label{dos_ss}
 \rho_{n\s s}^{W}(\omega)=\frac{1}{2(2\pi)^{2}nA_{n}^{2/n}v_{z}}\frac{\sqrt{\pi}\Gamma(1/n)}{\Gamma[(n+2)/2n]}
 |\omega+\mu+sQ_{0}|^{2/n},
\end{equation}
where we added a subscript ``$n$'' to highlight the dependence on the topological charge.
Thus, the total multi-Weyl density of states with both broken TRS and IS is given by
\begin{eqnarray}\label{multi-dos}
 \rho_{n}^{W}(\omega)&=&\sum_{\s s}\rho_{\s s}^{W}(\omega).
 \end{eqnarray}
We notice from Eqs.~\eqref{dos_ss} and~\eqref{multi-dos} that 
if both TRS and IS are preserved~(multi-Dirac system) or if only TRS symmetry is broken~(broken-TRS multi-Weyl system), then the density of states does not depend on either spin or chirality indices, and we have $\rho_{n}^{D}(\omega)=\rho_{nBTRS}^{W}(\omega)\equiv\rho_{n}(\omega)$ with 
\begin{eqnarray}\label{multi-dirac-dos}
\rho_{n}(\omega)=\frac{2}{(2\pi)^{2}nA_{n}^{2/n}v_{z}}\frac{\sqrt{\pi}\Gamma(1/n)}{\Gamma[(n+2)/2n]}
 |\omega+\mu|^{2/n}.
 \end{eqnarray}
In particular, for $n=1,2,3$, the multi-Weyl densities of states~\eqref{dos_ss} are 
\begin{eqnarray}
    \rho_{1\s s}^{W}(\omega)&=&\frac{1}{4\pi^{2}v_{\perp}^{2}v_{z}}
 |\omega+\mu+sQ_{0}|^{2}, \\
 \rho_{2\s s}^{W}(\omega)&=&\frac{1}{16\pi (k_{0}^{-1}v_{\perp})v_{z}}
 |\omega+\mu+sQ_{0}|, \\
 \rho_{3\s s}^{W}(\omega)&=&\frac{1}{24\pi^{2}(k_{0}^{-2}v_{\perp})^{2/3}v_{z}}\frac{\Gamma(1/3)}{\Gamma[5/6]}
 |\omega+\mu+sQ_{0}|^{2/3}.  \nonumber \\
\end{eqnarray}

\section{Mean-Field equations for the Multi-Dirac/Weyl Kondo lattice}
\label{sec:kondo_temperature}

In the multi-Dirac case with particle-hole symmetry we have only one mean-field parameter $V_{s}=V$.
Moreover, at $T=0$  $f(E^{++}(\veck))=f(E^{-+}(\veck))=0$, and $f(E^{+-}(\veck))=f(E^{--}(\veck))=1$, yielding
\begin{eqnarray}\label{eq_D}
 \int \frac{d^{3}\veck}{(2\pi)^{3}}\frac{1}{\sqrt{\epsilon_{k}^{2}+4V^{2}}}=\frac{1}{2J},
\end{eqnarray}
where $\epsilon_{k}=\sqrt{A_{n}^{2}k_{\perp}^{2n}+v_{z}^{2}k_{z}^{2}}$. 
Introducing the high-momentum cutoff $\Lambda$ and performing the change of variables~\eqref{changes} of Appendix~\ref{sec:weyl_dos}, we obtain the dimensionless equation
\begin{equation}\label{dirac_gap_1}
 \int_{0}^{1}dx\frac{x^{2/n}}{\sqrt{x^{2}+m^{2}}}=\frac{1}{\rho_{n}(\Lambda)J},
\end{equation}
where $m=2V/\Lambda$ and $\rho_n(\Lambda)$, given by~\eqref{multi-dirac-dos}, is here computed at $\Lambda$ with $\mu=0$. 
Eq.~\eqref{dirac_gap_1} has a non-trivial solution only above a critical Kondo coupling $J_{n}^{c}$, determined by letting $m=0$ in Eq.~\eqref{dirac_gap_1}, resulting in
\begin{eqnarray}
 J_{n}^{c}=\frac{(2/n)}{\rho_{n}(\Lambda)},
\end{eqnarray}
which is consistent with the  Withoff-Fradkin result for the pseudogap Kondo problem with $r=2/n$~\cite{PhysRevLett.64.1835}. 

In the absence of particle-hole symmetry~($\mu\neq 0$, $\lambda\neq 0$), using the same procedure as before, 
at $T=0$ $f(E^{++}({\veck}))=f(E^{+-}({\veck}))=f(E^{-+}({\veck}))=0$, and we obtain the self-consistency equations
\begin{eqnarray}
 \int_{0}^{1}dx\frac{x^{(2/n)+1}}{\sqrt{(x-\alpha)^{2}+m^{2}}}&=&\frac{1}{\rho_{n}(\Lambda)\Lambda}\left[1-\frac{\alpha}{J/\Lambda}\right] \label{sf1}, \\
  \int_{0}^{1}dx\frac{x^{2/n}}{\sqrt{(x-\alpha)^{2}+m^{2}}}&=&\frac{1}{\rho_{n}(\Lambda)J}\label{sf2}.
 \end{eqnarray}
where $\alpha=(\mu+\lambda)/\Lambda$.
 
When the TRS is broken but inversion symmetry is preserved, we have $Q\neq0$ and $Q_{0}=0$. Following along the same lines as in Appendix~\ref{sec:weyl_dos}, we can shift $k_{z}-sQ\rightarrow k_{z}$ in the mean-field equations and, for weak TRS breaking $\Lambda \gg Q$, the dependence on the TRS breaking parameter $Q$ disappears, and we conclude that $V_{s}=V$ for the broken-TRS multi-Weyl Kondo system. Thus, the broken-TRS multi-Weyl Kondo mean-field equations are identical to those of the multi-Dirac system~[Eqs.~\eqref{sf1}-\eqref{sf2}].

For the broken-IS multi-Weyl Kondo lattice, 
minimization of the mean-field free energy leads to the self-consistency equations
 \begin{eqnarray}
 &&\int_{0}^{1}dx \frac{x^{2/n}}{\sqrt{(x-\alpha-\bar{Q}_{0})^{2}+m_{+}^{2}}}=\frac{1}{\rho_{n}(\Lambda)J}, \label{TRS1}\\
 &&\int_{0}^{1}dx \frac{x^{2/n}}{\sqrt{(x-\alpha+\bar{Q}_{0})^{2}+m_{-}^{2}}}=\frac{1}{\rho_{n}(\Lambda)J}, \label{TRS3} \\
&&\int_{0}^{1}dx \frac{x^{(2/n)+1}}{\sqrt{(x-\alpha-\bar{Q}_{0})^{2}+m_{+}^{2}}} \nonumber \\
&&+\int_{0}^{1}dx \frac{x^{(2/n)+1}}{\sqrt{(x-\alpha+\bar{Q}_{0})^{2}+m_{-}^{2}}} \nonumber \\
&&=\frac{2}{\rho_{n}(\Lambda)\Lambda}\left[1-\frac{\alpha}{J/\Lambda}\right], \label{TRS2}
\end{eqnarray}
where $\alpha=(\mu+\lambda)/\Lambda$, $\tilde{Q}_{0}=Q_{0}/\Lambda$ and $m_{\pm}=2V_{\pm}/\Lambda$. In the limit $Q_{0}=0$ we recover Eqs.~\eqref{sf1}-\eqref{sf2}, as expected.

\section{Detailed calculations of the multi-Dirac Kondo quantum critical behavior}
\label{sec:critical}

In the vicinity of the quantum phase 
transition point $j_{n}=1$ ($J=J_{n}^{c}$) the Kondo 
coupling can be written in terms of a small 
perturbation $\delta$ as $j_{n}=1+\delta$, with $\delta\ll 1$. Using this in Eq.~\eqref{gap_dirac_f} and expanding up to linear order in $\delta$ we get
\begin{eqnarray}\label{fmd}
 f_{n}(m)=-\frac{n}{2}\delta,
\end{eqnarray}
where 
$$f_{n}(m)=\int_{0}^{1}dxx^{2/n}\left[\frac{1}{\sqrt{x^{2}+m^{2}}}-\frac{1}{x}\right].$$
Performing the variable change $x=my$ 
\begin{equation}\label{fm}
 f_{n}(m)=m^{2/n}\int_{0}^{1/m}dyy^{2/n}\left[\frac{1}{\sqrt{y^{2}+1}}-\frac{1}{y}\right].
\end{equation}
For $n>1$, we can let $m\rightarrow 0$ since the integral converges. In this case, 
\begin{equation}\label{critical_n}
 m(\delta)=C_{n}\delta^{n/2}, \qquad (n>1)
\end{equation}
where
$C_{n}=\left[\frac{-n\sqrt{\pi}}{\Gamma\left(\frac{1}{2}+\frac{1}{n}\right)\Gamma\left(-\frac{1}{n}\right)}\right]^{n/2}$. Thus, for $n>1$, the mean-field critical exponent is  $\nu=n/2$.

%

For $n=1$, we split the relevant integral as
\begin{eqnarray*}
 &&\int_{0}^{1/m}dyy^{2}\left[\frac{1}{\sqrt{y^{2}+1}}-\frac{1}{y}\right]=\int_{0}^{1}dyy^{2}\left[\frac{1}{\sqrt{y^{2}+1}}-\frac{1}{y}\right] \nonumber \\
 &&+
 \int_{1}^{1/m}dy\left\{y^{2}\left[\frac{1}{\sqrt{y^{2}+1}}-\frac{1}{y}\right]+\frac{1}{2y}\right\} -\frac{1}{2}\int_{1}^{1/m}\frac{dy}{y}.
\end{eqnarray*}
We can safely let $m\to 0$ in the second term on right-hand side since that integral is now convergent.  Using this in Eq.~\eqref{fmd} results in
\begin{equation}\label{critical_nn1}
m^{2}{\rm ln}\left(\frac{m}{A}\right)=-\delta,
\end{equation}
where $A=\sqrt{e}/2$.
Eq.~\eqref{critical_nn1} can be solved in the limit $\delta \ll 1$ yielding
\begin{equation}\label{cc}
m(\delta)=f(\delta)\delta^{1/2}, \qquad (n=1),
\end{equation}
where $f(\delta)=\sqrt{\frac{{-2}}{W_{-1}[-A^{2}\delta]}}$.
Here, $W_{-1}(x)$ is one of the branches of the Lambert function \cite{Corless1996}, which has the following asymptotic behavior for $0<x\ll 1$
\begin{eqnarray}\label{asympt}
W_{-1}(-x)&=&
\ln{(x)} - \ln{[-\ln{(x)}]
} + \smallO(1).
\end{eqnarray}
Thus, the critical exponent is  $\nu=1/2$, with logarithmic corrections built into the function~$f(\delta)$. 

\section{Analytical results for the multi-Dirac and broken-TRS multi-Weyl Kondo lattice }
\label{sec:analytical}

We can extract analytical results from the self-consistency Eqs.~\eqref{sf1}-\eqref{sf2} in the $n=1,2$ cases. To do so, we use the expressions 
\begin{eqnarray*}
&&\int_{0}^{1}dx\frac{x}{\sqrt{(x-\alpha)^{2}+m^{2}}}=\sqrt{m^{2}+(1-\alpha)^{2}}-\sqrt{m^{2}+\alpha^{2}} \nonumber \\
&&+\alpha\left[{\rm ln}\left(1-\alpha+\sqrt{m^{2}+(1-\alpha)^{2}}\right)\right. \\
&&\left.-{\rm ln}\left(\sqrt{m^{2}+\alpha^{2}}-\alpha\right)\right], \nonumber \\
\end{eqnarray*}
\begin{eqnarray*}
 &&\int_{0}^{1}dx\frac{x^{2}}{\sqrt{(x-\alpha)^{2}+m^{2}}}=\frac{1}{2}\left[\sqrt{m^{2}+(1-\alpha)^{2}} \right.\\
 &&\left.+3\alpha\left(\sqrt{m^{2}+(1-\alpha)^{2}}-\sqrt{m^{2}+\alpha^{2}}\right)\right] \\
 &&-\frac{1}{2}(m^{2}-2\alpha^{2})\left[{\rm ln}\left(1-\alpha+\sqrt{m^{2}+(1-\alpha)^{2}}\right)\right.\\
&&\left. -{\rm ln}\left(\sqrt{m^{2}+\alpha^{2}}-\alpha\right)\right],
\end{eqnarray*}
\begin{eqnarray*}
 &&\int_{0}^{1}dx\frac{x^{3}}{\sqrt{(x-\alpha)^{2}+m^{2}}}=\frac{1}{6\sqrt{(m^{2}+\alpha^{2})[m^{2}+(1-\alpha)^{2}]}}\times \\
 &&\left\{2\sqrt{m^{2}+\alpha^{2}}
 +4m^{4}\left(\sqrt{m^{2}+(1-\alpha)^{2}}-\sqrt{m^{2}+\alpha^{2}}\right)\right. \\
 &&\left.+ m^{2}\left[-2\sqrt{m^{2}+\alpha^{2}}+13\alpha\sqrt{m^{2}+\alpha^{2}}\right.\right. \\
 &&\left.\left. +7\alpha^{2}\left(\sqrt{m^{2}+\alpha^{2}}-\sqrt{m^{2}+(1-\alpha)^{2}}\right)\right] \right. \\
 &&\left.\alpha\left[\sqrt{m^{2}+\alpha^{2}}+3\alpha\sqrt{m^{2}+\alpha^{2}}-17\alpha^{2}\sqrt{m^{2}+\alpha^{2}} \right.\right.\\
 &&\left.\left.+11\alpha^{3}\left(\sqrt{m^{2}+\alpha^{2}}-\sqrt{m^{2}+(1-\alpha)^{2}}\right)\right]
 \right\}\\
 &&+\frac{\alpha}{2}(-3m^{2}+2\alpha^{2})\left[{\rm ln}\left(1-\alpha+\sqrt{m^{2}+(1-\alpha)^{2}}\right)\right.\\
 &&\left.-{\rm ln}\left(\sqrt{m^{2}+\alpha^{2}}-\alpha\right)\right].
\end{eqnarray*}
In the regime in which the Kondo parameter $m$ is exponentially small and $m\ll \alpha$, these equations simplify considerably. Then, if $n=1$, 
\begin{eqnarray}
 &&-\alpha^{2}{\rm ln}\left(\frac{m^{2}}{2\alpha}\right)+f_{1}(\alpha)=\frac{1}{\rho_{1}(\Lambda)J}, \\
 &&\frac{2\alpha}{\rho_{1}(\Lambda)J}+\frac{1}{3}(1-2\alpha^{3})=\frac{1}{\rho_{1}(\Lambda)\Lambda},
\end{eqnarray}
where $f_{1}(\alpha)=\frac{1}{2}\{1+2\alpha-6\alpha^{2}+2\alpha^{2}{\rm ln}[2(1-\alpha)]\}$. For $n=2$ we obtain
\begin{eqnarray}
 &&-\alpha{\rm ln}\left(\frac{m^{2}}{2\alpha}\right)+f_{2}(\alpha)=\frac{1}{\rho_{2}(\Lambda)J}, \\
 &&\frac{2\alpha}{\rho_{2}(\Lambda)J}+\frac{1}{2}(1-2\alpha^{2})=\frac{1}{\rho_{2}(\Lambda)\Lambda},
\end{eqnarray}
where $f_{2}(\alpha)=1-2\alpha+\alpha{\rm ln}[2(1-\alpha)]$. Closed-form expressions can be obtained when $0<\alpha\ll 1$, with exponential accuracy for $m$. For $n=1$,
\begin{eqnarray}
 \alpha&\approx&\frac{1}{2}\left(\frac{1}{\rho_{1}(\Lambda)\Lambda}-\frac{1}{3}\right)\rho_{1}(\Lambda)J, \\
 m&\approx&{\rm exp}\left[-\frac{1}{2\rho_{1}(\alpha)J}\right]. \nonumber \\
\end{eqnarray}
Analogously for $n=2$, 
\begin{eqnarray}
 \alpha&\approx&\left(\frac{1}{\rho_{2}(\Lambda)\Lambda}-\frac{1}{2}\right)\rho_{2}(\Lambda)J, \\
 m&\approx&{\rm exp}\left[-\frac{1}{2\rho_{2}(\alpha)J}\right]. \nonumber \\
\end{eqnarray}

\section{General Hamiltonian at the Particle-Hole Symmetric Point}\label{sec:symmetric}

At the particle-hole symmetric point~($\lambda=\mu=0$) and in the  presence of TRS~($Q=0$), some choices of the coupling 
parameters $J$, $W$, and $K$ allow for an analytical treatment, which we now describe.
In these cases, the dispersion relations of Hamiltonian~\eqref{H_HS} are given by 
\begin{eqnarray}\label{dispersion_general_1}
 E_{s}^{\alpha\xi}(\veck)=\frac{1}{2}\alpha\left[\sqrt{\e_{ks}^{2}+4r_{1}^{2}}+\xi\sqrt{\e_{ks}^{2}+4r_{2}^{2}}\right],
\end{eqnarray}
where $\alpha=\pm$, $\xi=\pm $, $\varepsilon_{ks}=\sqrt{A_{n}^{2}k_{\perp}^{2n}+v_{z}^{2}k_{z}^{2}}-s Q_{0}$, and $r_{1}$ and $r_{2}$ 
are order parameters which depend on the case in question. 

The first interesting case is $K=0$, for which $r_{1}=Jv$, $r_{2}=Ww$. By minimization of the free energy at $T=0$ if $J>W$ we have
\begin{equation}\label{gap_s}
 \sum_{s=\pm} \int_{0}^{1}dx\frac{x^{2/n}}{\sqrt{(x-sQ_{0})^{2}+m^{2}}}=\frac{2}{\rho_{n}(\Lambda)J},
\end{equation}
where $m=2Jv/\Lambda$,  
$\rho_{n}(\Lambda)$ is the multi-Weyl density of states~(see Appendix~\ref{sec:kondo_temperature}) and $w=0$~($r_{2}=0$). When $W>J$, we should exchange $(J,v)$ and $(W,w)$. In both cases, the dispersion relations are those of Eq.~\eqref{bands} with $\mu=\lambda=Q=0$, $V=Jv$ ~($J>W$) or $V=Ww$~($W>J$).

The second case is $J=W\neq K$, for which $r_{1}=Jv+Kw$ and $r_{2}=Kv+Jw$. After the free energy minimization we obtain $v/J=w/K$ and again the problem reduces to a single order parameter ruled by Eq.~\eqref{gap_s}. However, the dispersion relation 
structure is slightly different from Eq.~\eqref{bands} since $r_{1}$ and $r_{2}$ are both finite.


\section{Multi-Weyl Kondo semimetal specific heat}\label{sec:specific}

To compute the multi-Weyl Kondo semimetal specific heat we will follow Hsin-Hua Lai~\textit{et al.} in the study of the single Kondo-Weyl semimetal~\cite{doi:10.1073/pnas.1715851115}. We will take the effective multi-Weyl dispersion $\e_{ks}=\hbar \sqrt{A_{n}^{*2}k_{\perp}^{2n}+v_{z}^{*2}(k_{z}-sQ)^{2}}$, where we restore the fundamental constant $\hbar$,  $A_{n}^{*}=k_{0}^{(1-n)}v_{\perp}^{*}$, $v_{\perp}^{*}$ and $v_{z}^{*}$ are the renormalized velocities of the heavy Fermi liquid. The free-energy can be computed as 
\begin{eqnarray}
 F=-\frac{N}{\beta}\sum_{s=\pm}\int\frac{d^{3}\veck}{(2\pi)^{3}}{\rm ln}[1+e^{-\beta\e_{ks}}].
\end{eqnarray}
Performing the momentum integral above using the same variable transformations~\eqref{changes}, and using
\begin{eqnarray*}
 C_{v}(T)=-T\frac{\partial^{2} F}{\partial T^{2}},
\end{eqnarray*}
we obtain the specific heat for the multi-Weyl Kondo semimetal
\begin{equation}
 \frac{C_{v}(T)}{T}=\gamma_{n}T^{2/n},
\end{equation}
where
\begin{eqnarray}\label{gn}
 \gamma_{n}&=&\frac{k_{B}}{\pi^2}\left(\frac{k_{B}}{\hbar}\right)^{(2+n)/n}\frac{(1+n)(2+n)}{n^{4}} [\Gamma(1/n)]^2 \nonumber \\
&&\quad\times \zeta\left(2+\frac{2}{n}\right)\frac{[2^{(2+n)/n}-1]}{k_{0}^{n(1-n)/n}v_{\perp}^{*2/n}v^{*}_{z}},
\end{eqnarray}
and $\zeta(s)$ is the Riemann zeta function. 


This result agrees with reference~\cite{doi:10.1073/pnas.1715851115} in the particular case of $n=1$. In general, we see that the specific 
heat is enhanced relative to the non-interacting value by a factor of 
$(v_{\perp}/v_{\perp}^{*})^{2/n}(v_{z}/v_{z}^{*})$. 

%

\end{document}